\documentclass[amsmath,amssymb,twocolumn,aps,prd,nofootinbib,letterpaper,superscriptaddress,longbibliography,notitlepage]{revtex4-1}

\usepackage[usenames,dvipsnames]{color}

\usepackage{graphicx}

\usepackage{mathtools}

\usepackage{interval}

\usepackage{dblfloatfix}
\usepackage{csquotes}

\usepackage[inter-unit-separator=\cdot]{siunitx} % unités SI

\usepackage[caption=false]{subfig}
\usepackage{commath}
\usepackage{microtype}
\usepackage[english]{babel}
\usepackage[bookmarks]{hyperref}

\renewcommand{\Re}{\operatorname{Re}}
\renewcommand{\Im}{\operatorname{Im}}

\begin{document}

\title{Excited hairy black holes:\\dynamical construction and level transitions}
\author{Pablo Bosch}
\email{pbosch@perimeterinstitute.ca}
\affiliation{Perimeter Institute for Theoretical Physics, 31 Caroline Street North, Waterloo, Ontario, N2L 2Y5, Canada}
\affiliation{Department of Physics and Astronomy, University of Waterloo,
Waterloo, ON, N2L 3G1, Canada}
\author{Stephen R. Green}
\email{stephen.green@aei.mpg.de}
\affiliation{Max Planck Institute for Gravitational Physics (Albert Einstein Institute), Am M\"uhlenberg 1, 14476 Potsdam, Germany}
\author{Luis Lehner}
\email{llehner@perimeterinstitute.ca}
\affiliation{Perimeter Institute for Theoretical Physics, 31 Caroline Street North, Waterloo, Ontario, N2L 2Y5, Canada}
\author{Hugo Roussille}
\email{hugo.roussille@apc.in2p3.fr}
\affiliation{Laboratoire AstroParticule et Cosmologie (APC), UMR CNRS 7164, Université Paris Diderot, 10 Rue Alice Domon et Léonie Duquet 75013 Paris, France}

\date{\today}

\begin{abstract}
  We study the dynamics of unstable Reissner-Nordstr\"om anti-de
  Sitter black holes under charged scalar field perturbations in
  spherical symmetry. We unravel their general behavior and approach
  to the final equilibrium state. In the first part of this work, we
  present a numerical analysis of massive charged scalar field
  quasinormal modes. We identify the known mode
  families---superradiant modes, zero-damped modes, AdS modes, and the
  near-horizon mode---and we track their migration under variation of
  the black hole and field parameters. We show that the zero-damped
  modes become superradiantly unstable for large RNAdS with large
  gauge coupling; the leading unstable mode is identified with the
  near-horizon condensation instability. In the second part, we
  present results of numerical simulations of perturbed large RNAdS,
  showing the nonlinear development of these unstable modes. For
  generic initial conditions, charge and mass are transferred from the
  black hole to the scalar field, until an equilibrium solution with a
  scalar condensate is reached. We use results from the linear
  analysis, however, to select special initial data corresponding to
  an unstable overtone mode. We find that these data evolve to produce
  a new equilibrium state---an \emph{excited hairy black hole} with
  the scalar condensate in an overtone configuration. This state is,
  however, unstable, and the black hole eventually decays to the
  generic end state. Nevertheless, this demonstrates the potential relevance 
  of overtone modes as transients in black hole dynamics.
  
\end{abstract}

\maketitle

\section{Introduction}

Explorations of spacetime dynamics in general relativity have
uncovered many surprising phenomena with theoretical and astrophysical
implications. Examples include the discovery of critical
phenomena~\cite{Choptuik:1992jv,Gundlach:1999cu}, spacetime
turbulence~\cite{Bizon:2011gg,Carrasco:2012nf,Yang:2014tla,Adams:2013vsa},
and the black hole superradiant instability~\cite{Press:1972zz}, the
latter of which has been proposed as a probe of dark
matter~\cite{Arvanitaki:2010sy,East:2017mrj}. In recent years, the
AdS/CFT correspondence has provided additional motivation for studying
black holes in anti-de Sitter (AdS) spacetimes: black hole
equilibration is believed to be holographically dual to thermalization
of strongly coupled field theories, whereas instabilities describe
phase transitions~\cite{Hartnoll:2008kx,Gubser:2008px}.

One interesting theme is the explosion of black hole solutions when
standard assumptions are relaxed. Black holes with ``hair'' (i.e.,
stationary black holes described by quantities other than the total
mass, angular momentum, and electric charge) are generally forbidden
as asymptotically-flat solutions to the Einstein-Maxwell system in
four dimensions. But with
additional fields, higher dimensions, or more general boundary
conditions, the various theorems can be circumvented, and additional
solutions with the same conserved quantities can emerge~\cite{PhysRevD.51.R6608,Herdeiro:2015waa}. For instance,
in five dimensions, with one compactified dimension, there exist black
string \emph{and} black hole solutions. Generally, one of these
solutions will be entropically preferred, and this often implies
dynamical instability of the other
solutions~\cite{Hollands:2012sf}. Indeed, if the compactified
dimension is large compared to the black hole radius, then the black
string is linearly unstable~\cite{Gregory:1993vy}. Nonlinearly, the
string bifurcates self-similarly into a chain of black
holes~\cite{Lehner:2010pn}.

Black holes can have hair made up of additional fields if there is a
confining mechanism to prevent dissipation. This occurs, for instance,
in asymptotically AdS spacetimes, or for massive fields. One example
is a charged planar AdS black hole in the presence of a charged scalar
field: for sufficiently low temperature, there exist two stationary
solutions, Reissner-Nordstr\"om-AdS (RNAdS) and a charged black hole
with a scalar condensate. At these temperatures, RNAdS is subject to
the near-horizon scalar condensation instability~\cite{Gubser:2008px},
which leads to the hairy black hole under dynamical
evolution~\cite{Murata:2010dx}. For small RNAdS, the superradiant
instability also leads to a hairy black hole~\cite{Bosch:2016vcp}.

The hairy black holes obtained as end states of evolution
in~\cite{Murata:2010dx,Bosch:2016vcp} are in their \emph{ground
  state.} In the superradiant case, the final black hole can be
understood as an equilibrium combination of a small RNAdS black hole
with the fundamental mode of a charged scalar field in global
AdS~\cite{Dias:2016pma}. However, the scalar field also has overtone
solutions, and it is intriguing to ask whether these might also give
rise to hairy black holes, now in their \emph{excited state.} 

The central result of this paper is the dynamical construction of
stationary {\em excited hairy black holes}. Our approach is to start
with a fine-tuned perturbation of an unstable black hole that
corresponds to an unstable overtone quasinormal mode. We evolve the
instability numerically, and it eventually forms the excited hairy
black hole. This black hole, is, however, unstable, and after some
time decays to the ground state.

We take our initial black hole to be RNAdS, which is dynamically
unstable to charged scalars even in spherical symmetry. Although our
end goal is the excited hairy black hole, we begin in
section~\ref{sec:linear} with a numerical study of RNAdS massive
charged scalar field quasinormal modes. Ultimately, we use the results
of this analysis to construct the special initial data, but this
section also constitutes a thorough analysis of the various modes of
RNAdS throughout parameter space. Instabilities of RNAdS are
usually studied using approximations that rely on the smallness of
some parameter, either the black hole radius in the case of the
superradiant instability, or the surface gravity for the near-horizon
instability. We use the continued fraction method of
Leaver~\cite{Leaver:1985}, so our numerical analysis does not require
these approximations.

Previous analyses have identified several mode families. For small
black holes, RNAdS is ``close'' to global AdS, and therefore its
spectrum contains quasinormal modes that are deformations of AdS
normal modes. The normal-mode frequencies are evenly spaced along the
real axis, so for sufficiently large gauge coupling $q$, they can be
made to satisfy the superradiance condition,
$0 < \Re \omega < q Q / r_+$. Modes satisfying this condition are
amplified when they interact with the black hole, leading to
instability~\cite{Uchikata:2011zz}.

Extremal RNAdS, meanwhile, has a near-horizon region with metric
$\text{AdS}_2 \times S^2$~\cite{Bardeen:1999px}. This gives rise to an
instability whenever the effective near-horizon mass of the scalar
field lies below the Breitenlohner-Freedman (BF)
bound~\cite{Breitenlohner:1982bm} of the near-horizon region and the
true mass is kept above the global BF bound~\cite{Gubser:2008px}. This
condition is most easily satisfied for large black holes in global
AdS~\cite{Dias:2016pma}, and by continuity it also extends to
near-extremal black
holes~\cite{PhysRevD.82.124033,PhysRevD.81.124020,Hollands:2014lra}.

In addition to the AdS modes, which can be superradiantly unstable,
and the near-horizon mode, the spectrum of RNAdS also contains a
collection of ``zero-damped'' modes. These modes are associated to the
near-horizon region of near-extreme black holes, and indeed they are
present also in the asymptotically flat
case~\cite{Zimmerman:2016qtn}. For small RNAdS they are described by a
tower of evenly-spaced quasinormal frequencies extending below the
real axis near the superradiant-bound frequency, with imaginary part
proportional to the surface gravity. As extremality is approached,
these merge into a branch point representing the horizon instability
of Aretakis~\cite{Aretakis:2012ei}.

The interplay between superradiant and near-horizon instabilities was
studied in~\cite{Dias:2016pma}, where it was shown that for small
black holes, the near-horizon instability condition for a massless
field becomes $q^2 > 1/(4 r_+^2)$, so that it ceases to operate for
fixed $q$ as the black hole is made smaller. Conversely, the
near-horizon instability does not require superradiance, as it will
occur with $q=0$ provided $m^2$ is sufficiently
negative~\cite{PhysRevD.81.124020}.

All of the modes above can be seen in our figures in
section~\ref{sec:linear-results}. Our numerical results, however,
provide further clarity on the nature of the near-horizon
instability. We show by varying the black hole size and the gauge
coupling, that the zero-damped modes for small black holes migrate to
become superradiantly unstable for large black holes. This family of
modes has in fact many similarities to the small black hole AdS
modes. Finally, we show that the leading unstable mode migrates to
become the near-horizon unstable mode under suitable variation of
parameters.

Although a dynamical instability can be identified through a
linearized analysis, this cannot capture its complete time
development. In section~\ref{sec:nonlinear} we present results of
numerical simulations showing the full nonlinear development of the
unstable modes of large RNAdS in spherical symmetry. In
section~\ref{sec:generic}, we evolve generic initial perturbations,
and observe a dynamical behavior similar to that observed
in~\cite{Bosch:2016vcp} for the small RNAdS superradiant instability:
the modes extract charge and mass until the system settles to a final
static black hole with a scalar condensate. For smaller gauge
coupling, the final condensate lies closer to the black hole, similar
to simulations of the near-horizon instability in the planar
limit~\cite{Murata:2010dx}.

In section~\ref{sec:excited}, we construct the excited hairy black
holes. We select parameters such that the corresponding large RNAdS
solution has more than one unstable mode. Taking the quasinormal modes
from the analysis of section~\ref{sec:linear}, we carefully perturb
the background RNAdS solution with the first overtone mode, $n=1$. We
observe that under evolution the field extracts charge and mass until
the mode saturates and superradiance stops. This time, however, the
black hole is in an $n=1$ excited state. This black hole appears to be
a stationary solution, but it is in fact unstable to the fundamental
$n=0$ perturbation, since only the $n=1$ mode saturated the
superradiant bound. Because of small nonlinearities and numerical
errors, we cannot avoid seeding this mode, albeit at much smaller
amplitude. After some time, it grows and overtakes the $n=1$
mode, and the black hole decays to the ground state.

As a final demonstration, we construct initial data consisting of
\emph{several} superradiant modes, such that the solution cascades
through a series of unstable excited black hole equilibria
corresponding to different overtones. A sample evolution is shown in
figure~\ref{fig:cascade}.
\begin{figure}[tb]
  \centering
  \includegraphics[width=\linewidth]{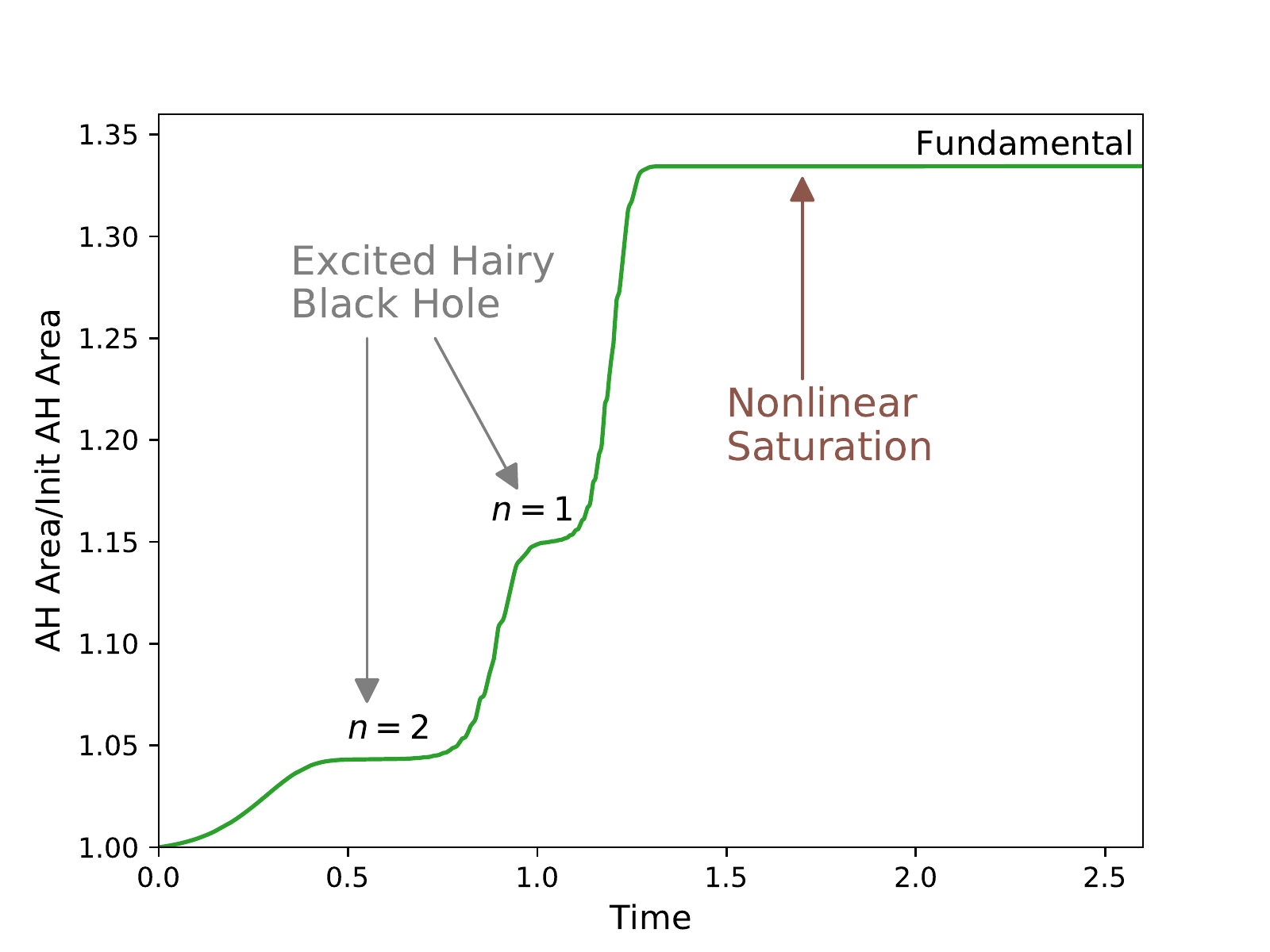}
  \caption{Apparent horizon area of a black hole as it undergoes a
    series of transitions through metastable excited hairy black hole
    states. Initial data chosen to consist primarily of $n=2$
    overtone, with subleading $n=1$ overtone.}
  \label{fig:cascade}
\end{figure}

This paper is organized as follows. In section~\ref{sec:model} we
introduce the Einstein-Maxwell-charged scalar system and the RNAdS
background solution. In section~\ref{sec:linear}, we review the
various mode families of RNAdS and we present our linear analysis. We
present our nonlinear simulations of the instabilities in
section~\ref{sec:nonlinear}, with the construction of the excited
hairy black holes in section~\ref{sec:excited}. We conclude in
section~\ref{sec:conclusion}. Throughout the paper, unless otherwise
indicated we follow conventions of~\cite{Wald:1984} and we work in
four spacetime dimensions.

\section{Model}\label{sec:model}

We consider Einstein gravity with a negative cosmological constant,
coupled to Maxwell and massive charged scalar fields. The Lagrangian
density is
\begin{equation}
\mathcal{L} = R + \frac{6}{L^2} - F_{ab}F^{ab} -
4 \left(\left | D_a \psi \right|^2 - m^2|\psi|^2\right) ,
\label{eq:Lagrangian}
\end{equation}
where $D_a \equiv \nabla_a - iqA_a$ is the gauge covariant
derivative. This gives rise to the Einstein equation,
\begin{equation}
G_{ab} - \frac{3}{L^2}g_{ab} = 8\pi T_{ab}^{\psi} + 8\pi T_{ab}^{\mathrm{EM}},
\label{eq:Einstein}
\end{equation}
with stress-energy tensors,
\begin{align}
  T_{ab}^{\psi} ={}& \frac{1}{4\pi}\left\{ \frac{1}{2}\left[ (D_a \psi)^\ast (D_b \psi) + (D_a \psi) (D_b \psi)^\ast \right] \right. \nonumber\\
                &\left.-\frac{1}{2}g_{ab}\left(|D_c\psi|^2 + m^2|\psi|^2\right) \right\} , \label{eq:StressPsi}\\
  T_{ab}^{\mathrm{EM}} ={}& \frac{1}{4\pi}\left ( g^{cd}F_{ac}F_{bd} - \frac{1}{4}
g_{ab} F_{cd} F^{cd} \right ) \label{eq:StressEM},
\end{align}
the Maxwell equation,
\begin{equation}
\nabla^b(\nabla_b A_a - \nabla_a A_b ) = iq\psi^\ast D_a \psi -
iq\psi(D_a \psi)^\ast ,
\label{eq:Maxwell}
\end{equation}
and the Klein-Gordon equation,
\begin{equation}
D^a D_a \psi - m^2\psi  = 0.
\label{eq:ScalarField}
\end{equation}

The RNAdS black hole is a static, spherically symmetric solution with
vanishing $\psi$. In Boyer-Lindquist coordinates, the metric is
\begin{equation}\label{eq:rnads-g}
  ds^2 = -f(r) dt^2 + \frac{1}{f(r)} dr^2 + r^2d\Omega_2^2,
\end{equation}
with
\begin{equation}
  f(r)= 1- \frac{2M}{r} + \frac{Q^2}{r^2} + \frac{r^2}{L^2},
\end{equation}
and the Maxwell potential is
\begin{equation}\label{eq:rnads-A}
  A_{\mu} dx^{\mu} = \left(-\frac{Q}{r} + C\right) dt.
\end{equation}
We have inserted a constant $C dt$ in $A$, which is pure gauge. We
will take $C=0$ in most of this paper, so that $A(\infty)=0$. In
section~\ref{sec:NH-instability}, however, we will take $C=Q/r_+$ to
set $A(r_+)=0$, which is convenient for studying the near-horizon
geometry. Under a change of gauge $C \to C + \Delta C$, the scalar
field undergoes a frequency shift $\psi \to e^{iq\Delta Ct}\psi$.

We take the RNAdS solution \eqref{eq:rnads-g}--\eqref{eq:rnads-A}
as the background for the linear analysis. Also imposing spherical
symmetry, the Klein-Gordon equation~\eqref{eq:ScalarField} takes the
form
\begin{align}\label{eq:KG-expanded}
  0 ={}& -\frac{1}{f}\partial_t^2\psi +\frac{2iqA_0}{f}\partial_t\psi + \frac{1}{r^2}\partial_r\left(r^2f\partial_r\psi\right)  \nonumber\\
  & -\left(m^2 - \frac{q^2A_0^2}{f}\right)\psi.
\end{align}
Since $\psi$ vanishes in the background, it decouples from the other
fields at linear order, and it is consistent to study $\psi$ as a test
field. We analyze~\eqref{eq:KG-expanded} in section~\ref{sec:linear}.

It is convenient to express the background quantities in terms of the
inner and outer horizon radii, $r_-$ and $r_+$. The metric function
becomes
\begin{align}
  f(r) ={}& \frac{(r-r_-)(r-r_+)}{L^2r^2}\left[ r^2 + \left(r_-+r_+\right)r \right. \nonumber\\
  &\left.+ \left(L^2 + r_-^2 + r_+^2 + r_-r_+\right) \right],
\end{align}
from which we can read off the mass and charge of the black hole,
\begin{align}
  M &= \frac{r_-+r_+}{2L^2}\left[ \left(r_+-r_-\right)^2 + L^2 \right],\label{eq:mass-BH}\\
  Q^2 &= \frac{r_-r_+}{L^2} \left(L^2 + r_-^2 + r_+^2 + r_-r_+ \right).\label{eq:charge-BH}
\end{align}
Thus, at extremality, $r_+ = r_-$, and
\begin{align}
  M_{\text{ext}} &= r_+,\\
  Q^2_{\text{ext}} &= r_+^2\left(1 + \frac{3 r_+^2}{L^2}\right).
\end{align}

In section~\ref{sec:nonlinear}, we solve the full system of equations
\eqref{eq:Einstein}--\eqref{eq:ScalarField} with $m=0$ in spherical
symmetry in ingoing coordinates.

\section{Linear perturbations}\label{sec:linear}

In this section we study the test scalar
field~\eqref{eq:KG-expanded}. We begin in
section~\ref{sec:linear-prelim} by describing the mode families and
instabilities that we expect to see in our numerics. In
section~\ref{sec:linear-formalism} we describe the continued fraction
method for finding quasinormal frequencies numerically, and we present
our results in section~\ref{sec:linear-results}.

\subsection{Preliminaries}\label{sec:linear-prelim}

This section describes three families of modes that appear in the
spectra we obtain in section~\ref{sec:linear-results}: the
near-horizon mode, the AdS modes, and the zero-damped modes. These
have all been derived analytically under various approximations
elsewhere in the literature. We include them for completeness and for
interpreting our numerical results in
section~\ref{sec:linear-results}.

\subsubsection{Near-horizon condensation instability}
\label{sec:NH-instability}

Four-dimensional extremal black holes with spherical horizon topology
have near-horizon geometries closely related to
$\mathrm{AdS}_2 \times S^2$~\cite{Bardeen:1999px}; for extremal RNAdS,
this correspondence becomes exact. The near-horizon instability is
based on the violation of the BF bound of the near-horizon geometry by
the scalar field. Holographically, the condensation corresponds to a
transition to a superconducting phase below a critical
temperature~\cite{Gubser:2008px,Hartnoll:2008kx}.

To take the near-horizon limit it is useful to set the constant
$C=Q/r_+$ in the Maxwell field, so that $A_a$ vanishes on the
horizon. For extremal RNAdS, we then have
\begin{align}
  ds^2_{\text{ext}} &= -f_{\text{ext}}dt^2 + \frac{1}{f_{\text{ext}}}dr^2 + r^2d\Omega_2^2,\\
  A_{\text{ext}} &= \frac{Q_{\text{ext}}(r-r_+)}{rr_+}dt,
\end{align}
where
\begin{equation}
  f_{\text{ext}} = \frac{(r-r_+)^2}{L^2r^2}\left[ r^2 + 2r_+r + (L^2 + 3r_+^2)\right].
\end{equation}

We then define a change of coordinates depending on a parameter
$\lambda>0$,
\begin{equation}
  t = \frac{\tilde{t}}{\lambda}, \qquad r = r_+ + \lambda \tilde{r}.
\end{equation}
Taking the $\lambda\to0$ limit in these coordinates, we obtain the
near-horizon fields,
\begin{eqnarray}
  ds^2_{\text{NH}} &=& -\frac{\tilde{r}^2}{R^2} d\tilde{t}^2 + \frac{R^2}{\tilde{r}^2} d\tilde{r}^2 + r_+^2 d\Omega_2^2, \label{eq:NearHorizonLineElement} \\
  A_{\text{NH}} &=& \frac{Q_{\text{ext}} \tilde{r}}{r_+^2}d\tilde{t},
\end{eqnarray}
where
\begin{equation}
  \frac{1}{R^2} = \frac{6}{L^2} + \frac{1}{r_+^2}.
  \label{eq:AdS2Radius}
\end{equation}
The metric \eqref{eq:NearHorizonLineElement} is recognized as
$\text{AdS}_2\times S^2$ in Poincar\'e coordinates, where the $\text{AdS}_2$
factor has radius $R$. Note that the choice of $C$ ensures that the
Maxwell field remains finite in the near-horizon limit.

The scalar field acquires an effective mass in the near-horizon
region. Taking the near-horizon limit of the Klein-Gordon
equation~\eqref{eq:KG-expanded}, this is seen to be
\begin{align}\label{eq:meffsquared}
  m_{\text{eff}}^2  &= m^2 - \left.\frac{q^2A_{0,\text{ext}}^2}{f_{\text{ext}}}\right|_{\lambda\to0}\nonumber\\
  &= m^2 - q^2 \cdot \frac{L^2 + 3 r_+^2}{L^2 + 6 r_+^2}.
\end{align}
In the large black hole limit, $m_{\text{eff}}^2 \to m^2 -
q^2/2$. Instability can occur if $m_{\text{eff}}^2$ lies below the
near-horizon BF bound,
\begin{equation}
\label{eq:NHBF}
m_{\text{NHBF}}^2 = -\frac{1}{4R^2} = -\frac{1}{4}\left( \frac{6}{L^2} + \frac{1}{r_+^2} \right),
\end{equation}
which in the large black hole limit becomes
$m_{\text{NHBF}}^2 \to -3/(2L^2)$. It was futher shown using energy
arguments that for large black holes this bound is
sharp~\cite{Dias:2010eu}. To be globally stable, it is necessary that
the global BF bound be respected, i.e., $m^2 \ge - 9 / (4L^2)$. Thus,
in the large black hole limit, the near-horizon instability is
triggered if
\begin{equation}\label{eq:largeNH}
  -\frac{9}{4L^2} \le m^2 < -\frac{3}{2L^2} + \frac{q^2}{2},
\end{equation}
which can be easily satisfied by choosing sufficiently large $q^2$ or
negative $m^2$ (but not too negative). By continuity, the instability
is expected to also occur for near-extreme black
holes~\cite{Hartnoll:2008kx,Hollands:2014lra}.

For small black holes, it is not possible to trigger the near-horizon
instability with negative $m^2$ since in this case the near-horizon BF
bound is below the global BF bound. In addition, $q^2$ must be taken
very large to obtain an instability, i.e.,
\begin{equation}\label{eq:small-nh}
  q^2 \ge \frac{1}{4r_+^2} + m^2.
\end{equation}
For these reasons, the near-horizon instability is said to not operate
for small black holes~\cite{Dias:2016pma}.

It should be noted that in the rest of the paper we will set the gauge
constant $C\to0$, so mode frequencies pick up an additional shift
$\Delta \omega = qC = qQ/r_+$. In that gauge, the near-horizon unstable mode
frequency for near-extreme black holes will lie near the superradiant
bound frequency, $qQ/r_+$.

\subsubsection{Superradiant instability}
\label{sec:superradiant-instability}

The superradiant instability (or ``black hole bomb'') occurs when
superradiant scattering is combined with a confinement mechanism, such
as a mirror, a mass term, or an AdS
boundary~\cite{Press:1972zz}. Under superradiant scattering, an
incident wave is amplified by the black hole as it extracts mass and
angular momentum or charge. With the confinement mechanism, the
outgoing wave cannot escape to infinity, and instead interacts
repeatedly with the hole, resulting in exponential growth.

The superradiant condition is most easily derived from thermodynamic
arguments~\cite{PhysRevD.7.949}. In the charged black hole case,
consider a mode solution $\psi = e^{-i\omega t}R(r)$ with real
frequency $\omega$. The charge to mass ratio of the mode is
\begin{equation}
\frac{\delta Q}{\delta M} = \frac{q}{\omega}.
\label{eq:BHModeInteraction}
\end{equation}
When the mode interacts with the black hole, it exchanges charge and
mass in this ratio.  The first law of black holes mechanics for
charged black holes, however, is
\begin{equation}
\delta M = \frac{\kappa}{8\pi} \delta A_{\mathrm{H}} - \Phi_{\mathrm{H}}\delta Q,
\label{eq:FirstLaw}
\end{equation}
where $\kappa$ is the surface gravity, $A_{\text{H}}$ is the horizon
area, and $\Phi_{\mathrm{H}}$ is the electrostatic potential at the
horizon. Inserting \eqref{eq:BHModeInteraction} into
\eqref{eq:FirstLaw} relates the change in mass to the change in area
of the black hole as a consequence of interacting with the mode,
\begin{equation}
\delta M = \frac{\kappa}{8\pi}
\frac{\omega}{(\omega + \Phi_{\mathrm{H}}q )} \delta
A_{\mathrm{H}} .
\label{eq:FirstSR}
\end{equation}

The second law of black holes mechanics states that the area of the
horizon can only increase in dynamical processes,
$\delta A_{\mathrm{H}}\geq 0$. Hence, waves that satisfy
\begin{equation}
 0 < \omega < - q \Phi_{\mathrm{H}}  = \frac{qQ}{r_+}
\label{eq:SRCondition}
\end{equation}
will have $\delta M <0$, and will therefore extract mass and charge
from the black hole.

The modes themselves are provided by the confinement mechanism. For
small RNAdS, there is a set of modes that are deformations of
global AdS normal modes, which have frequencies,
\begin{equation}
  \omega_n^{\pm} = \pm \frac{2n + 3}{L}, \qquad n=0,1,2,\ldots
\end{equation}
We therefore expect instability for $\omega_n^+$ with
\begin{equation}
2n \lesssim qQL/r_+ - 3.
\label{eq:crit-SR-smallBH}
\end{equation}
By choosing $q$ sufficiently large, this condition is easily
satisfied.

For more detailed derivations of the superradiant instability using
matched asymptotic expansions we refer the reader
to~\cite{Uchikata:2011zz,Dias:2016pma}.

\subsubsection{Zero-damped modes}\label{sec:zdm}

A final class of modes that is relevant to our analysis is associated
to the near-horizon region of near-extremal black holes. These
long-lived ``zero-damped'' modes can be viewed as trapped in the
extended black hole throat region.

In the asymptotically flat case, these modes were shown
in~\cite{Zimmerman:2016qtn} to fall into one of two families,
\emph{principal} or \emph{supplementary}, depending on the charge
coupling and angular mode number of the scalar field. (The terminology
refers to the near-horizon $SO(2,1)$ representations in which these
modes lie.)  Using a matched asymptotic expansion, the quasinormal
frequencies of asymptotically-flat RN in spherical symmetry can be
shown~\cite{Zimmerman:2016qtn} to be
\begin{equation}\label{eq:princ}
  \omega_n^{\text{P}} = \frac{qQ}{r_+} + \kappa\left[ qr_+ - i \left(\frac{1}{2} - \sqrt{\frac{1}{4} - (qr_+)^2} + n\right) + \eta r_+ \right]
\end{equation}
and
\begin{equation}\label{eq:supp}
  \omega_n^{\text{S}} = \frac{qQ}{r_+} + \kappa\left[ qr_+ - i \left(\frac{1}{2} + \sqrt{\frac{1}{4} - (qr_+)^2} + n\right)\right],
\end{equation}
where $n=0,1,2,\ldots$ is the overtone number,
$\kappa$ is the surface gravity, and $\eta$ is a
small complex number. If the quantity under the square root is
positive, then the supplementary family~\eqref{eq:supp} applies,
otherwise the principal family~\eqref{eq:princ}.

We see that both families consist of a $\kappa$-spaced tower of
modes extending below the superradiant bound frequency. As
$\kappa\to0$ these modes converge to $qQ/r_+$, which becomes a branch
point; this is associated to the horizon instability of
Aretakis~\cite{Zimmerman:2016qtn}.

Notice that the quantity under the square root in
\eqref{eq:princ}--\eqref{eq:supp} becomes negative when the
near-horizon instability condition~\eqref{eq:small-nh} is satisfied,
i.e., the effective mass violates the near-horizon BF bound. The
frequency~\eqref{eq:princ} nevertheless does not correspond to an
instability in asymptotically-flat RN, as the imaginary part remains
negative. For small RNAdS, we expect\footnote{We thank P. Zimmerman
  for helpful discussions on this point.} similar behavior, with small
$(r_+/L)$-corrections to the quasinormal frequencies. For large RNAdS,
however, we will show numerically in section~\ref{sec:linear-large}
that these modes can become unstable when the near-horizon BF bound is
violated.

\subsection{Continued fraction method}
\label{sec:linear-formalism}

We now describe the continued fraction method for finding quasinormal
mode solutions. We seek solutions that are ingoing at the horizon and
satisfy the reflecting condition at the AdS boundary. Satisfaction of
both of these conditions should yield a discrete spectrum of complex
frequencies.

Let $\psi = e^{-i\omega t}R(r)$ be our mode ansatz. The Klein-Gordon
equation~\eqref{eq:KG-expanded} reduces to the radial equation,
\begin{equation}
  \label{eq:field-RN-AdS}
  \frac{d}{dr} \left(r^2 f \frac{dR}{dr} \right) + \left( \frac{(\omega r^2 - q Q r)^2}{r^2 f} - m^2 r^2 \right) R = 0 .
\end{equation}
Asymptotically as $r\to r_+,\infty$, this has two solutions,
\begin{equation}\label{eq:cases}
  R(r) \sim
  \begin{cases}
    (r-r_+)^{\pm \frac{i L^2 r_+ \left(r_+ \omega - qQ\right)}{\left(r_+-r_-\right) \left(L^2+r_-^2+3 r_+^2+2 r_- r_+\right)}} & \text{as } r\to r_+,\\
    r^{\frac{1}{2} \left(\pm\sqrt{4 m^2 L^2+9}-3\right)} & \text{as }r\to \infty.
  \end{cases}
\end{equation}
At infinity, we require the solution to decay, so we take the solution
with the minus sign as $r\to\infty$ in~\eqref{eq:cases}. This
corresponds to a reflecting condition at the AdS boundary. To impose
the ingoing condition at the horizon, we take the minus sign solution
as $r\to r_+$.

To find a solution everywhere with the desired asymptotic behavior, we
write the radial function as
\begin{equation}
  R(r) = (r-r_+)^A (r-r_-)^{B-A} F(u),
\end{equation}
where 
\begin{align}
  A &= -\frac{i L^2 r_+ \left(r_+ \omega - qQ\right)}{\left(r_+-r_-\right) \left(L^2+r_-^2+3 r_+^2+2 r_- r_+\right)},\\
  B &= - \frac{1}{2} \left(\sqrt{4 m^2 L^2+9} + 3\right),
\end{align}
and where $F(u)$ is a new unknown function with $u =
(r-r_+)/(r-r_-)$. If $F$ is smooth on $u \in [0,1]$ then $R$ satisfies
the desired asymptotic conditions.

We now expand $F$ as a power series
\begin{equation}
\label{eq:powseries}
F(u) = \sum_{n=0}^{\infty} a_n u^n,
\end{equation}
and insert this into~\eqref{eq:field-RN-AdS}. This gives a complicated
relation on the sequence $(a_n)$, which we can simplify into a
three-term recurrence relation using Gaussian reduction. We finally
obtain a relation of the form
\begin{equation}
  \begin{aligned}
    \alpha_0 a_1 + \beta_0 a_0 &= 0,\\
    \alpha_n a_{n+1} + \beta_n a_n +\gamma_n a_{n-1} &= 0, \qquad  n \ge 1,
  \end{aligned}
\label{eq:rec-three-RN-AdS}
\end{equation}
where $\alpha_n$, $\beta_n$, and $\gamma_n$ all depend on $\omega$ and
the system parameters. We obtained complicated closed form expressions
for these coefficients, but have not included them due to space
considerations.

If the series~\eqref{eq:powseries} converges uniformly for some value
of $\omega$, then that corresponds to a quasinormal mode. To obtain
such $\omega$ we use the continued fraction method of
Leaver~\cite{Leaver:1985} and Gautschi~\cite{Gautschi:1967}. (This
method was recently used to compute quasinormal modes for a massless
charged scalar in asymptotically flat RN~\cite{Richartz:2014}.) The
method relies on the fact that the power series~\eqref{eq:powseries}
converges uniformly if the following continued fraction converges:
\begin{equation}
  \label{eq:contfrac}
  \frac{- \gamma_1}{\beta_1 - \frac{\alpha_1 \gamma_2}{\beta_2 - \frac{\alpha_2 \gamma_3}{\beta_3 - ...}}}.
\end{equation}
Moreover, if the continued fraction~\eqref{eq:contfrac} converges,
then it is equal to $a_1/a_0$. This allows us to close the
recurrence relation~\eqref{eq:rec-three-RN-AdS},
\begin{equation}
  \label{eq:main-contfrac}
  \frac{- \gamma_1}{\beta_1 - \frac{\alpha_1 \gamma_2}{\beta_2 - \frac{\alpha_2 \gamma_3}{\beta_3 - ...}}} = - \frac{\beta_0}{\alpha_0}.
\end{equation}

Quasinormal frequencies are the values of $\omega$ that
solve~\eqref{eq:main-contfrac}. We find these frequencies numerically
by plotting the logarithm of the difference between the left and right
sides of~\eqref{eq:main-contfrac}, and searching for
negative singularities.

\subsection{Results}\label{sec:linear-results}

We now present the mode spectra obtained numerically as we vary the
black hole parameters, $\alpha\equiv Q/Q_{\text{ext}}$ and $r_+$, and
the field parameters, $q$ and $m$. We fix $L=1$.

\subsubsection{Small black hole}

A typical quasinormal spectrum for small RNAdS is shown in
figure~\ref{fig:fracont_a_0.8_rp_0.1_L_1.0_q_4.0_mu_0.0_N_2500_l_0}. This
shows two branches of modes: the vertical branch extending below the
real axis is the supplementary branch~\eqref{eq:supp} of zero-damped
modes, and the horizontal branch is the family of AdS modes. The $n=0$
positive-frequency AdS mode lies within the band
$0 < \Re \omega < qQ/r_+$, and is therefore superradiantly unstable;
it lies above the real axis in
figure~\ref{fig:fracont_a_0.8_rp_0.1_L_1.0_q_4.0_mu_0.0_N_2500_l_0}.

\begin{figure}
  \includegraphics[width=\columnwidth]{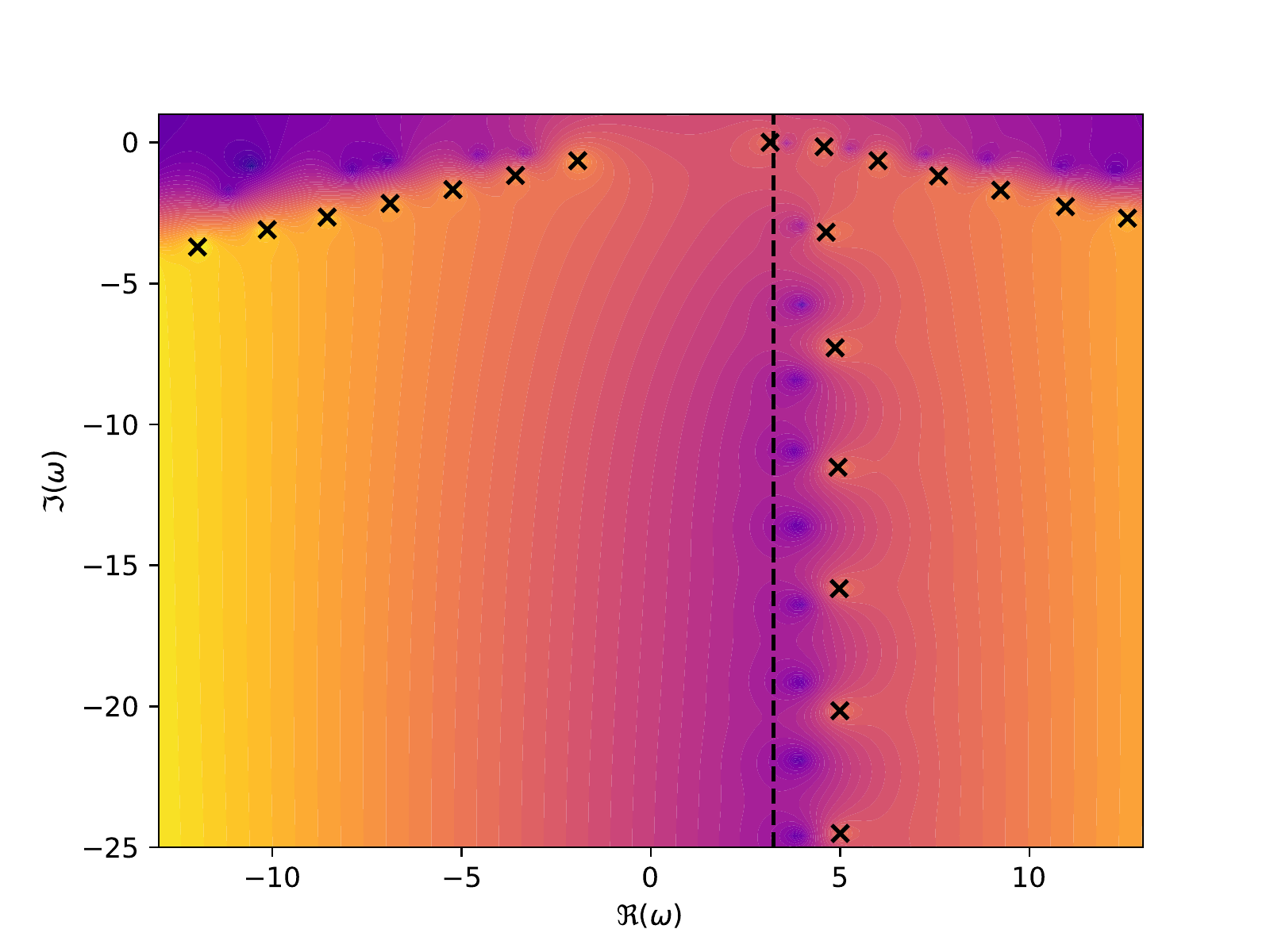}
  \caption{Continued fraction values for RNAdS with $r_+ = 0.1$,
    $\alpha = 0.8$, $q = 4$, and $m = 0$. We plot the logarithm of the difference between the left hand side and right hand side of~\eqref{eq:main-contfrac}. Darker colors 
    correspond to higher values. Quasinormal modes (minima in the plot) are marked by black
    crosses. We see that nonzero $q$ breaks the symmetry between
    positive and negative real part. Two branches of modes are
    present, a vertical branch of zero-damped modes, and a more
    horizontal branch AdS modes. The dashed vertical line corresponds to the
    superradiant bound frequency; modes satisfying
    $0 < \Re \omega < qQ/r_+$ are unstable.
    \label{fig:fracont_a_0.8_rp_0.1_L_1.0_q_4.0_mu_0.0_N_2500_l_0}}
\end{figure}

\begin{figure}[!h]
  \includegraphics[width=\columnwidth]{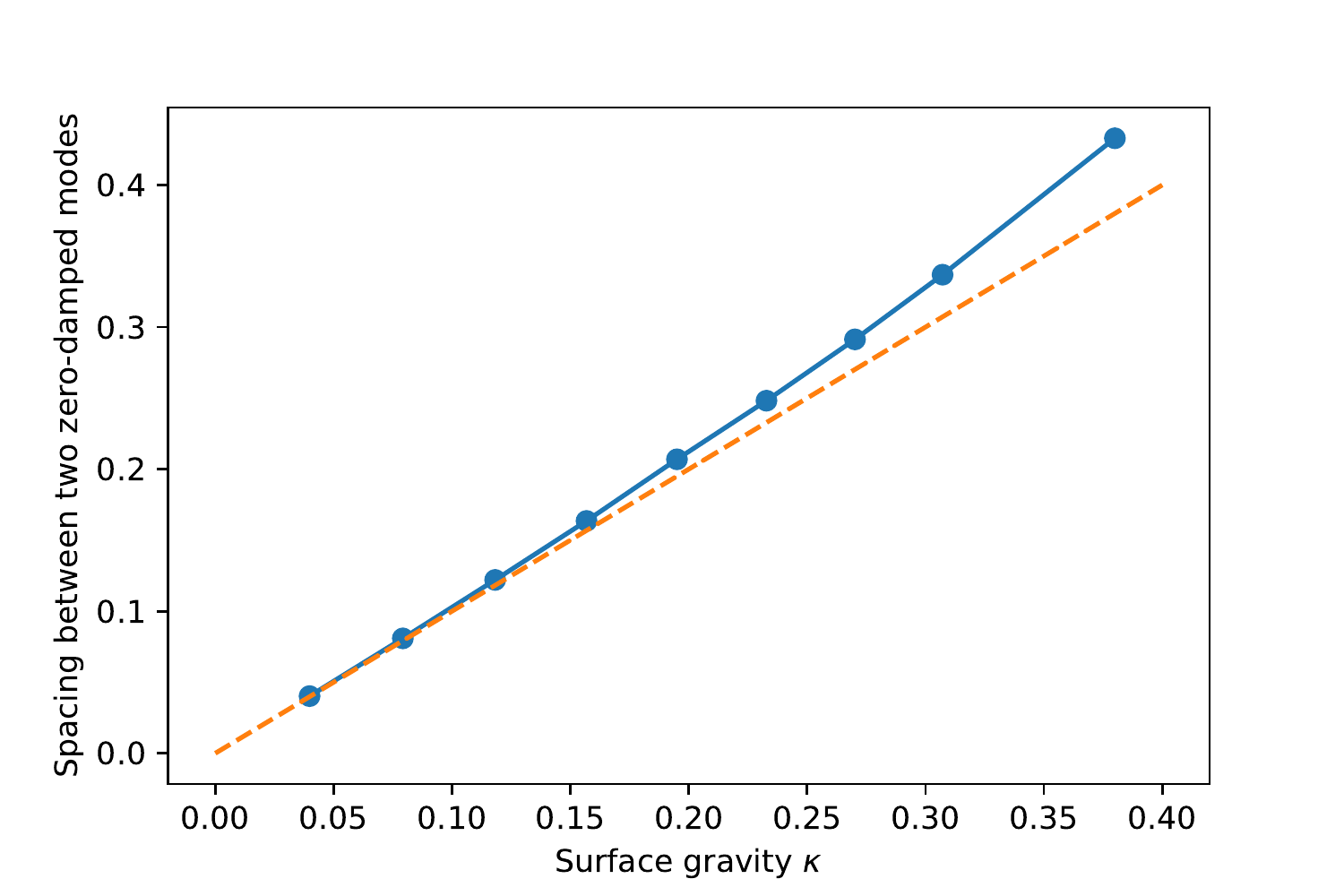}
  \caption{Spacing $\delta$ between two zero-damped modes versus
    surface gravity $\kappa$. Here, we take $q=0$ and $r_+ = 1$. We
    see that $\delta \to \kappa$ as $\kappa\to0$. \label{fig:ZDMs}}
\end{figure}

\begin{figure}[!h]
  \includegraphics[width=\columnwidth]{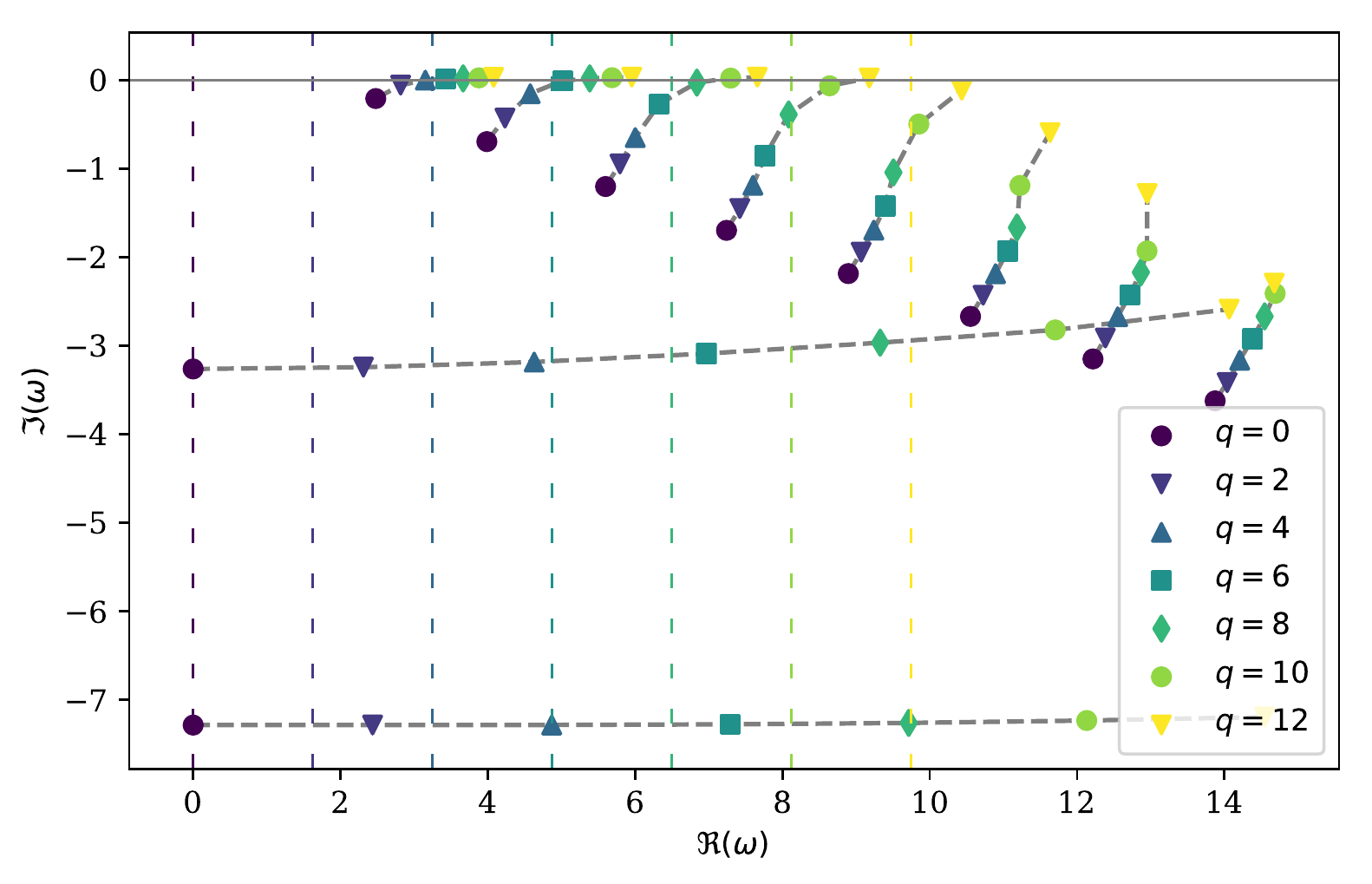}
  \caption{Quasinormal frequencies of RNAdS for $r_+ = 0.1$,
    $\alpha=0.8$, $m=0$, and $0 \le q \le 12$. AdS modes become unstable when $\Re \omega$
    drops below the superradiant bound frequency (indicated with dashed vertical lines).
    \label{fig:merging-RN-AdS}}
\end{figure}

For the stable AdS modes, we see that the decay rate is proportional
to the distance from the superradiant strip. We have also verified
that the separation between zero-damped modes is equal to $\kappa$:
this is shown in figure~\ref{fig:ZDMs} (This separation, although
expected to hold only for small RNAdS, holds also for larger $r_+$.) In this figure, we plot the mean spacing
$\delta = \langle \omega_{n+1}^{\text{S}} - \omega_n^{\text{S}}
\rangle$ for $r_+ = 1$ and $q=0$, as a function of the surface
gravity $\kappa$. We observe that $\delta \to \kappa$ as $\kappa\to0$,
in agreement with \cite{Dias:2016pma}.

We now study the influence of varying the field charge $q$ on the
spectrum; this is shown in figure~\ref{fig:merging-RN-AdS}. As $q$
increases, so does the superradiant bound frequency, $qQ/r_+$. The
tower of zero-damped mode frequencies remains tied to this frequency,
and shifts to the right in the complex plane as well. The AdS modes also
shift to the right, but more slowly than the superradiant bound
frequency. One by one, these modes are overtaken by the superradiant
bound frequency, and they become unstable. This is shown in figure~\ref{fig:merging-RN-AdS}.

\subsubsection{Large black hole}\label{sec:linear-large}

\begin{figure}[!h]
  \includegraphics[width=\columnwidth]{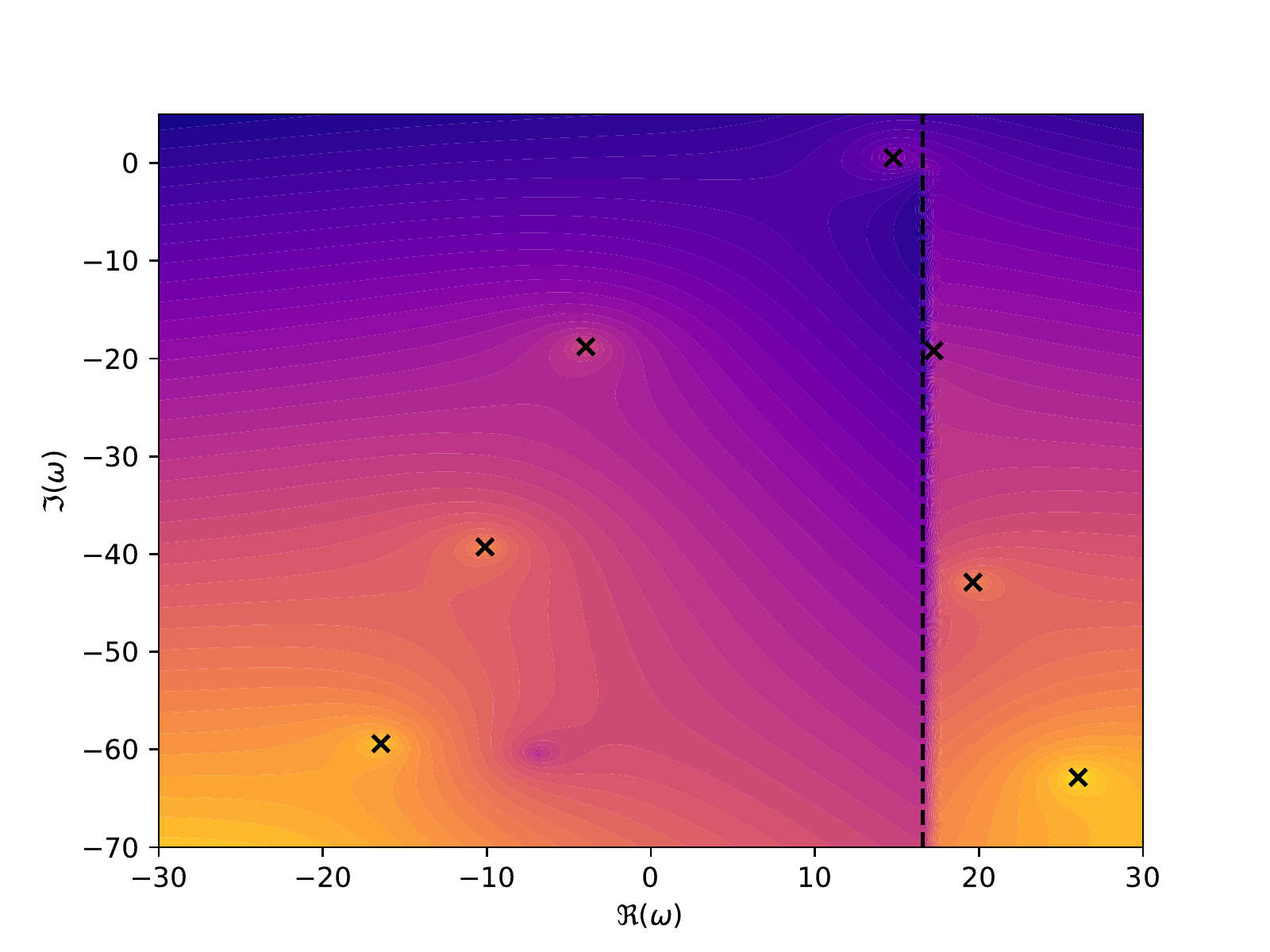}
  \caption{Continued fraction values for RNAdS with $r_+ = 5.0$ and $\alpha = 0.95$, and field parameters $q = 2$, $m^2 = -1.49^2$. We see two diagonal branches, and one unstable mode. Not visible due to resolution is the set of zero-damped modes.
    \label{fig:largebhspec}}
\end{figure}

\begin{figure}[!h]
\centering
\includegraphics[width=\columnwidth]{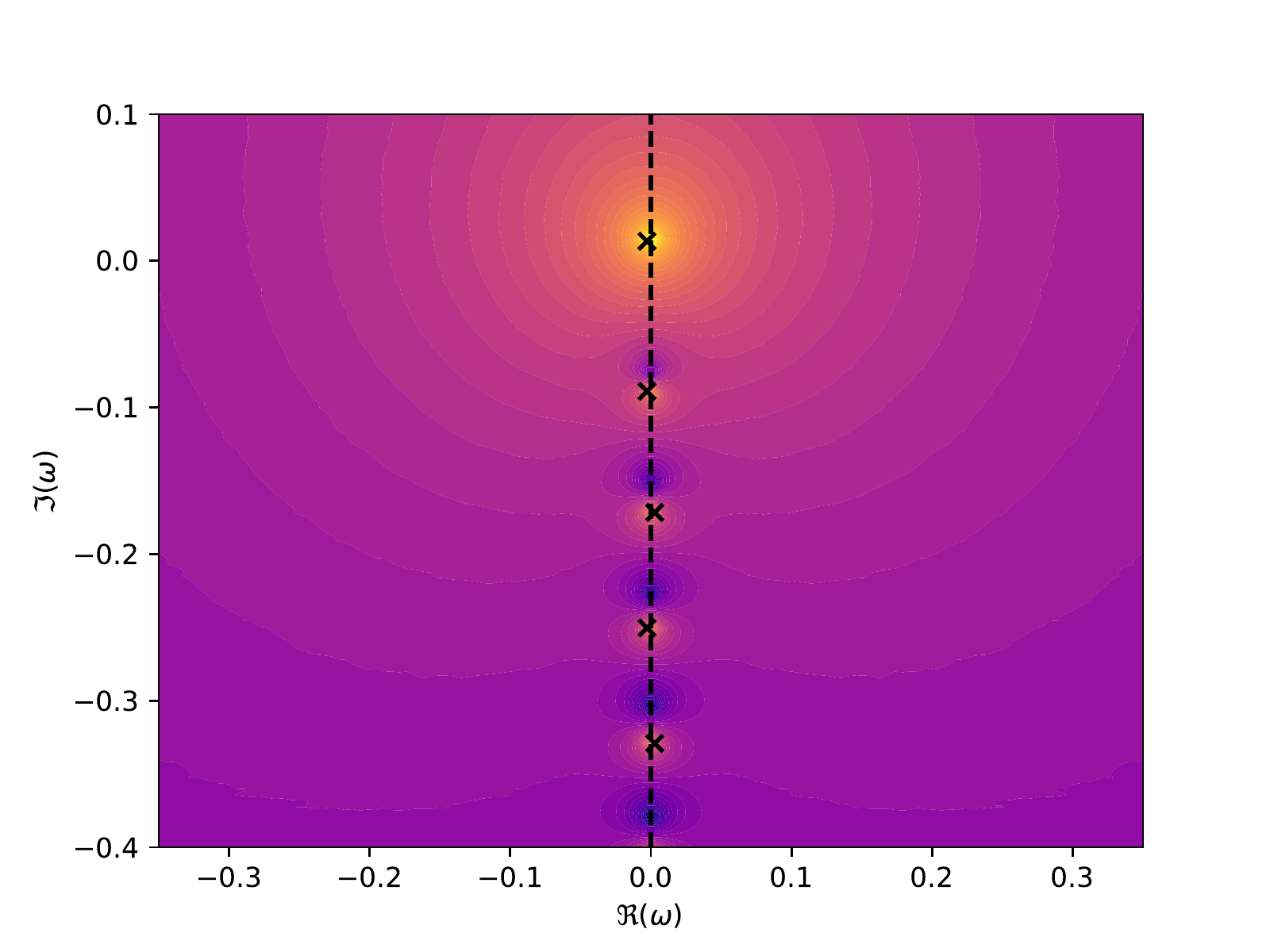}
\caption{Quasinormal modes for $\alpha=0.995$, $r_+=5$, $q=0$ and
  $m^2 = -1.49^2$. One mode in unstable here: this corresponds to the
  near-horizon mode.}
\label{fig:typical-plot-NH}
\end{figure}

The discussion of section~\ref{sec:linear-prelim} indicates that for
large RNAdS, we should see a near-horizon unstable mode and a tower of
zero-damped modes. AdS modes, meanwhile, are known to be present only
for small black holes, where they can be superradiantly unstable for
large $q$, and it is not clear in general what role superradiance
might play for large black holes. To disentangle the various
instabilities, we use our numerical code to find and track the
quasinormal modes.

A typical large RNAdS spectrum is shown in
figure~\ref{fig:largebhspec}. This shows two diagonal branches of
stable modes and one unstable mode. Already it is clear that the
superradiant instability plays a role for large black holes, as the
unstable frequency lies within the superradiant strip. In our studies,
we found that, when it exists, every unstable mode lies within this
strip.

To isolate the near-horizon instability, we can set $q=0$ to turn off
superradiance. In the extremal case, the tachyonic instability can
then be obtained with negative $m^2$ such that
\begin{equation}
  - \frac{9}{4L^2} \le m^2 < -\frac{3}{2L^2}.
\end{equation}
We choose $m^2 = -1.49^2$ to bring the mass squared close to the
global BF bound, and we consider a near-extremal black hole with
$\alpha=0.995$. For these parameters, the quasinormal frequencies are
plotted in figure~\ref{fig:typical-plot-NH}. This shows one unstable
mode, the near-horizon mode. The near-horizon mode lies close to the
real axis and is thus weakly unstable. It is also apparently isolated,
as we have not been able to identify a second near-horizon unstable
mode when $q=0$.

Figure~\ref{fig:typical-plot-NH} also shows a tower of stable modes
along the imaginary axis; these are the zero-damped modes. They are
evenly spaced, whereas the near-horizon mode is separated by a larger
distance. However, as we increased $m^2$ and decreased $\alpha$
to turn off the near-horizon instability, the modes re-positioned
themselves into a single family. Indeed, all modes shifted downward,
with the tachyonic mode dropping below the real axis and spacing
itself evenly at the top of the zero-damped family. Thus, the
near-horizon unstable mode is simply a member of the zero-damped
family of modes. Note that the zero-damped modes should also be
present in figure~\ref{fig:largebhspec}, but they are not clearly
visible due to lack of resolution in this figure, and the fact that
these modes are very closely spaced.

\begin{figure}[!h]
  \includegraphics[width=\columnwidth]{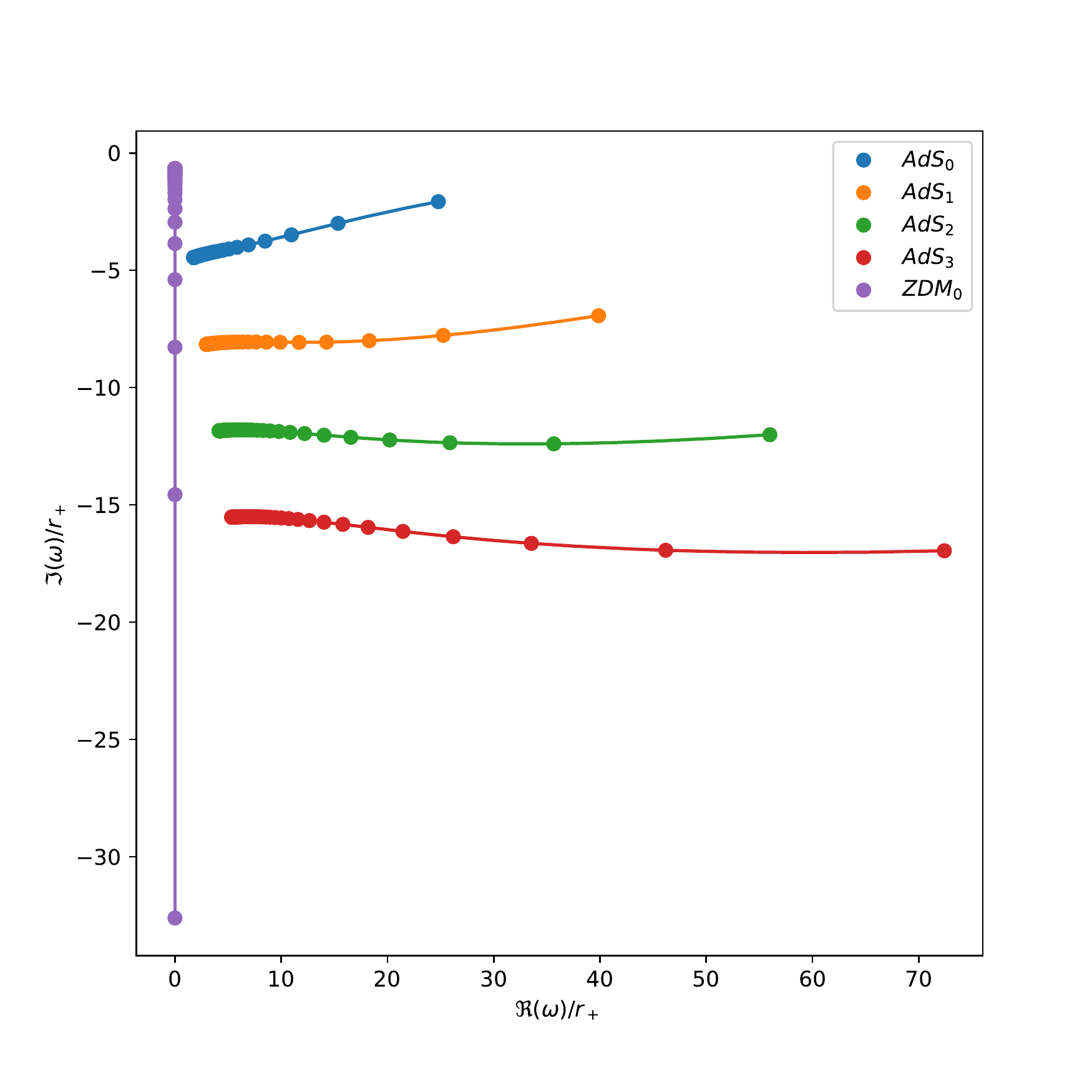}
  \caption{
Tracking of the first four AdS modes ($\text{AdS}_i$) and
    the first zero-damped mode ($\text{ZDM}_0$) for $r_+$ varied
    between 0.1 (where the points are far apart) and $5$ (where the
    points are close together) by steps of $0.05$, for $\alpha = 0.8$, $q=0$,
    and $m=0$. We observe a scaling for large
    black hole radius: when $r_+ \gg L$, the modes are invariant under
    the transformation $r_+ \rightarrow \lambda r_+$,
    $r_- \rightarrow \lambda r_-$,
    $\omega \rightarrow \lambda \omega$. This can be derived
    from~\eqref{eq:field-RN-AdS}. We also find that whereas for small
    RNAdS, the AdS modes are longest lived, for large RNAdS the
    zero-damped modes are longest lived.
    \label{fig:tracking-AdS-modes-rp}}
\end{figure}

We would now like to understand the connection between the modes of
large and small RNAdS. To do so, we first track the mode frequencies as the
size of the black hole is varied for
$q=0$. Figure~\ref{fig:tracking-AdS-modes-rp} shows the migration of
several AdS modes and the leading zero-damped mode as $r_+$ is varied
between $0.1$ and $5$. We observe that the diagonal branches of the
large black hole in figure~\ref{fig:largebhspec} correspond to the AdS
modes for small black holes. The importance of the different mode
families seems to be reversed for small and large RNAdS: for large
black holes, the zero-damped modes have slowest decay, whereas the AdS
modes are longest lived in the small black hole case.

\begin{figure}[!h]
	\includegraphics[width=\columnwidth]{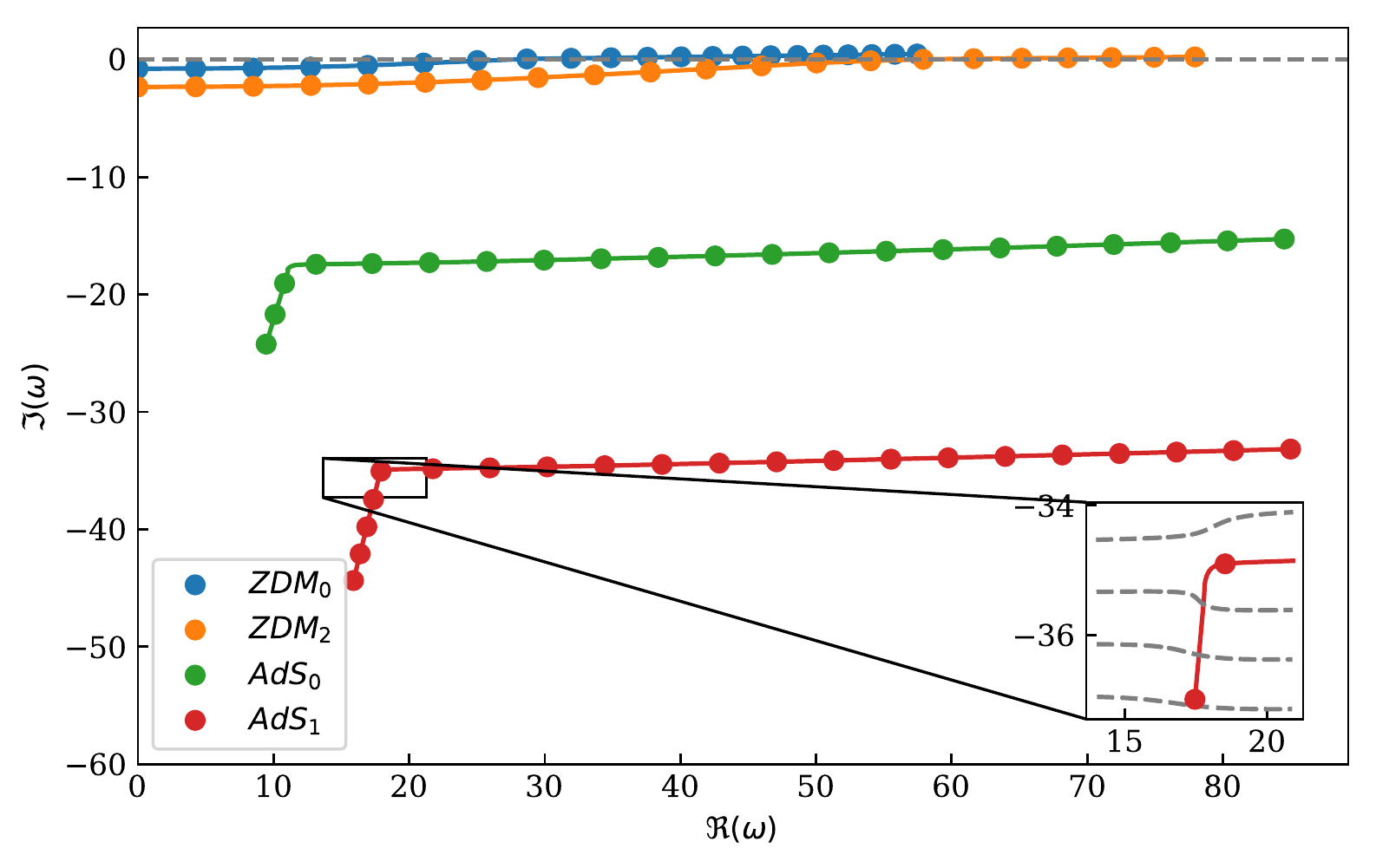}
	\caption{Quasinormal frequencies for varying $q$ for a large
          black hole with $r_+ = 0.1$, $\alpha=0.95$, and $m=0$. The
          gauge coupling $q$ varies from $0$ to $10$ by steps of
          $0.5$. We observe that the modes that become unstable are
          the zero-damped modes. The AdS mode frequencies pass through
          a kink in their migration: the inset shows how the second
          AdS mode slots itself in between two zero-damped
          modes.\label{fig:plot_tracking_q}}
\end{figure}

Next, we increase the gauge coupling $q$ to connect the near-horizon
mode to the general quasinormal spectrum of
figure~\ref{fig:largebhspec}; results are presented in
figure~\ref{fig:plot_tracking_q}. We observe a very different behavior
from the small black hole case of
figure~\ref{fig:merging-RN-AdS}. First, the modes that become unstable
are the \emph{zero-damped modes}, not the AdS modes. This is not
predicted by~\eqref{eq:supp}, which holds only for small black
holes. Once unstable, zero-damped modes have a spectrum similar to the
small black hole AdS-mode spectrum: the mode with smallest $\Re\omega$
has highest growth rate, and all unstable modes lie within the
superradiant strip. Second, the AdS mode frequencies pass through a
kink as they evolve; closer inspection reveals that they actually
merge into the tower of zero-damped modes at large $q$.

Thus, the mode corresponding to the near-horizon instability is also
the fastest growing mode in the case of a large black hole. For
$q > 0$, this mode lies within the superradiant strip, so violation of
the near-horizon BF bound and superradiance both contribute to its
instability. This is not the case for a small black hole, where the
fastest-growing mode is the first AdS mode, and instability occurs
even when the near-horizon BF bound is satisfied.

To summarize, for small RNAdS, unstable modes come from the AdS
branch, whereas for large RNAdS, they come from the zero-damped
branch. When the near-horizon BF bound is violated for large black
holes, the most unstable mode also exhibits near-horizon
instability. Figure~\ref{fig:tracking-AdS-modes-rp} shows the
crossover between the large and small black hole scenarios.

\section{Nonlinear evolution}\label{sec:nonlinear}

For our nonlinear studies, we solve the system of equations
\eqref{eq:Einstein}--\eqref{eq:ScalarField} numerically, with
$m=0$. As in the rest of the paper, we impose spherical symmetry and
reflecting boundary conditions at infinity.

In the following subsection we describe our numerical method. We then
describe the evolution for generic scalar field initial data in
subsection~\ref{sec:generic} and the excited hairy black hole in
subsection~\ref{sec:excited}.

\subsection{Method}

We use the same numerical code as we used
in~\cite{Bosch:2016vcp}. This uses ingoing Eddington-Finkelstein
coordinates $(v,r)$, similar to~\cite{Chesler:2013lia}, but adapted to
spherical symmetry. Equations are discretized with finite differences,
using mixed second and fourth order radial derivative operators
satisfying summation by parts (see,
e.g.,~\cite{Calabrese:2003vx,Calabrese:2003yd}) and fourth order
Runge-Kutta time stepping.

The spatial domain extends from an inner radius $r_0$, several grid
points within the apparent horizon, to infinity. The singularity is
thereby excised from the computational domain. To reach infinity, the
domain is compactified by working with a spatial coordinate
$\rho = 1/r$, and defining a uniform grid on the domain
$0 \leq \rho \leq 1/r_0$.

Boundary data consist of the ADM mass $M$ and charge $Q$, and initial
data are fully specified by the initial value of the scalar field,
$\psi(v=0)$. The system is solved by integrating radially inward along
$v=\text{constant}$ null curves to obtain the remaining field values
and their time derivatives; $\psi$ is then integrated one step forward
in time, and the procedure is iterated. With $\psi(v=0)=0$, this gives
RNAdS with mass $M$ and charge $Q$ as the solution, but more generally
some of the mass and charge is contained in the scalar field. The
characteristic formulation has some residual gauge freedom, which we
use to set the Maxwell potential to vanish at infinity, and to set $r$
to be the areal radius. (In~\cite{Chesler:2013lia} this is used to fix
the position of the horizon.) For further details, we refer the reader
to~\cite{Bosch:2016vcp,Bosch:2017ccw}.

The scalar field can be expanded about infinity, and with the
reflecting boundary condition, this takes the form,
\begin{equation}
  \psi(v,r) = \frac{\varphi_3(v)}{r^3} + O\left(\frac{1}{r^4}\right).
\end{equation}
The quantity $\varphi_3(v)$ is an output of the simulation, and it
contains information about the mode content of the solution. Other
gauge-invariant output quantities are the apparent horizon area
$A_{\text{H}}$ and the distribution of charge between the black hole
and the scalar field. To track the superradiant bound we extract the
electrostatic potential at the apparent horizon
$\Phi_{\mathrm{AH}}(v)$.

\subsection{Generic evolution}\label{sec:generic}

We now study the evolution of large RNAdS black holes perturbed with
generic scalar field configurations. We take the background solution
to have $r_+=100$ and $Q=0.9Q_{\text{ext}}$, fixing $L=1$
throughout. Strictly speaking, we only have control over the ADM
quantities (which we impose as boundary data) but we take the initial
scalar field to have very small amplitude, so to a good approximation
these directly determine the background black hole parameters.

We take the scalar field initial data to be compactly supported
outside the black hole, with profile
$\psi(v=0) = (r^{-1}-r_1^{-1})^3\ (r^{-1}-r_2^{-1})^3\ (\kappa_1 +
\kappa_2\sin(10r^{-1}))/r^2$ for $r_+ < r_1 \le r \le r_2 < \infty$,
and zero otherwise.  The observed dynamics are qualitatively similar
to the small black hole superradiant instability~\cite{Bosch:2016vcp}:
when the black hole is unstable, charge and mass are extracted by the
scalar field until a stationary hairy black hole final state is
reached. The final state is, moreover, independent of the initial
scalar field profile.

\begin{figure}[tb]
  \includegraphics[width=\linewidth]{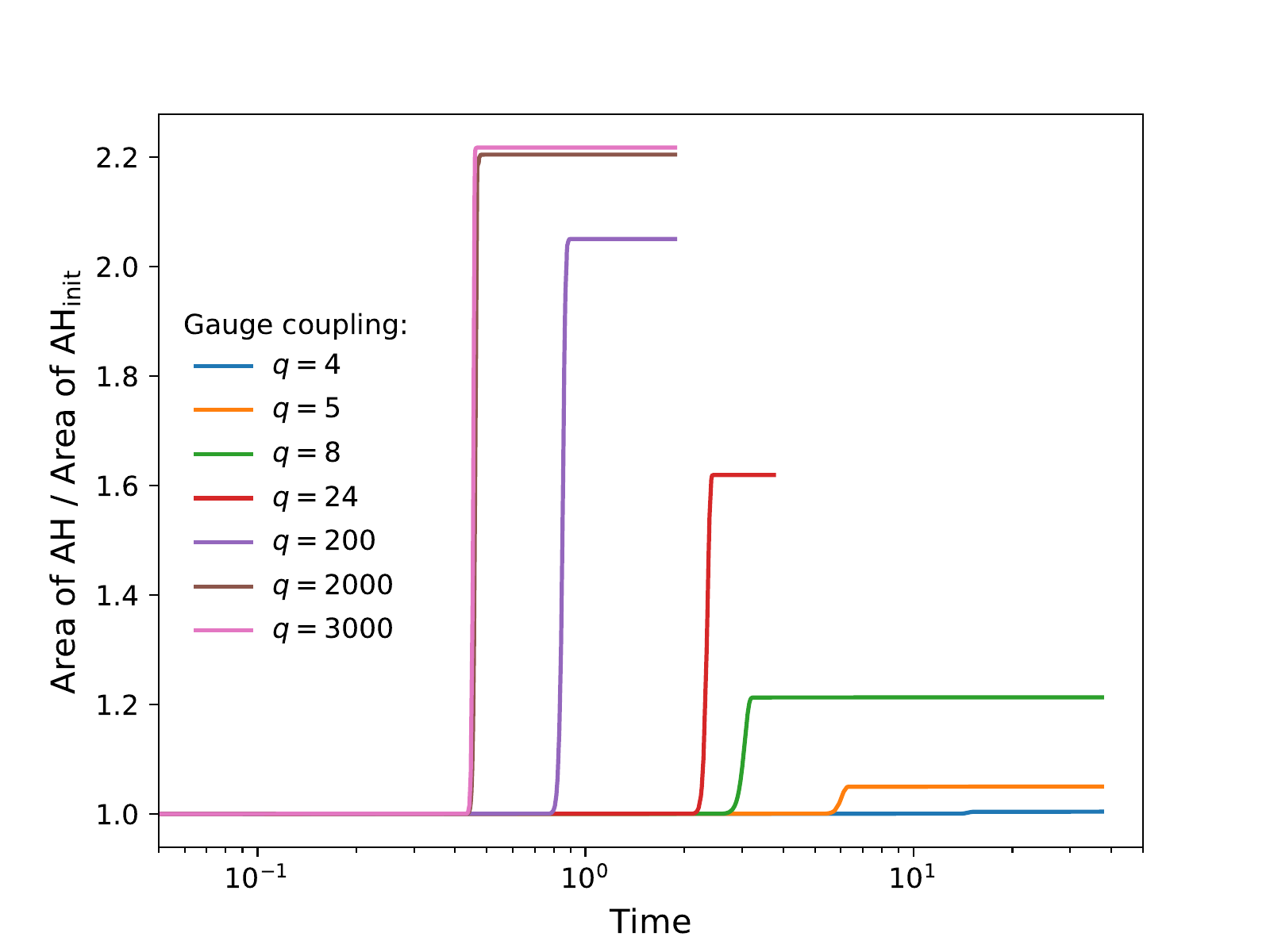}
  \caption{Normalized area of the apparent horizon as a function of
    time, for initial black hole with $r_+ = 100$ and
    $Q=0.9 Q_{\text{ext}}$. Different colors correspond to different
    gauge coupling $q$.\label{fig:AHarea_SR}}
\end{figure}

\begin{figure}[tb]
  \includegraphics[width=\linewidth]{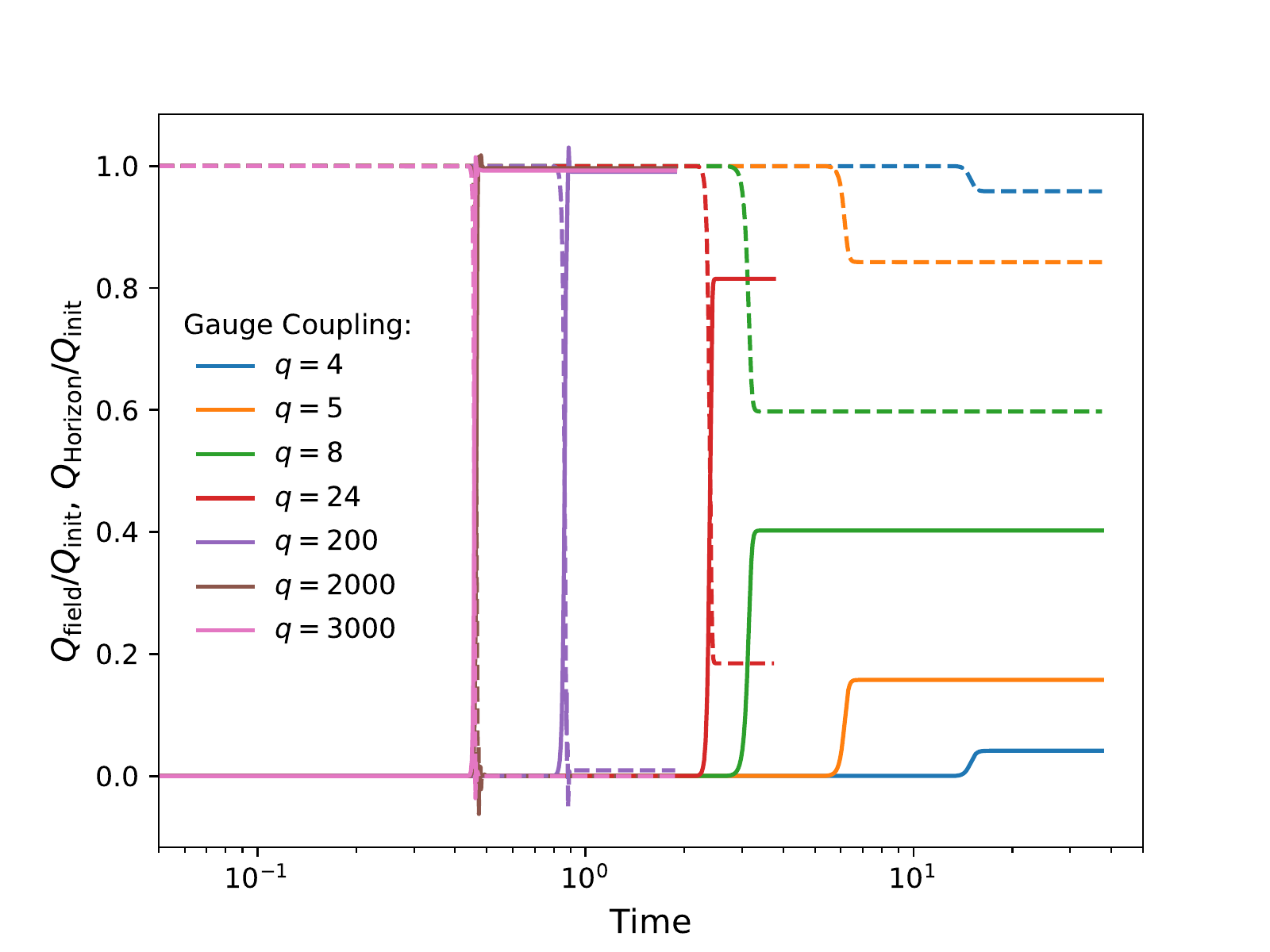}
  \caption{Normalized charge contained in the black hole and in the
    scalar field as a function of time, for initial black hole with
    $r_+ = 100$ and $Q=0.9Q_{\text{ext}}$. Solid lines denote scalar
    field charge, and the dashed lines denote black hole
    charge.\label{fig:charge_LSR}}
\end{figure}

\begin{figure}[tb]
  \includegraphics[width=\linewidth]{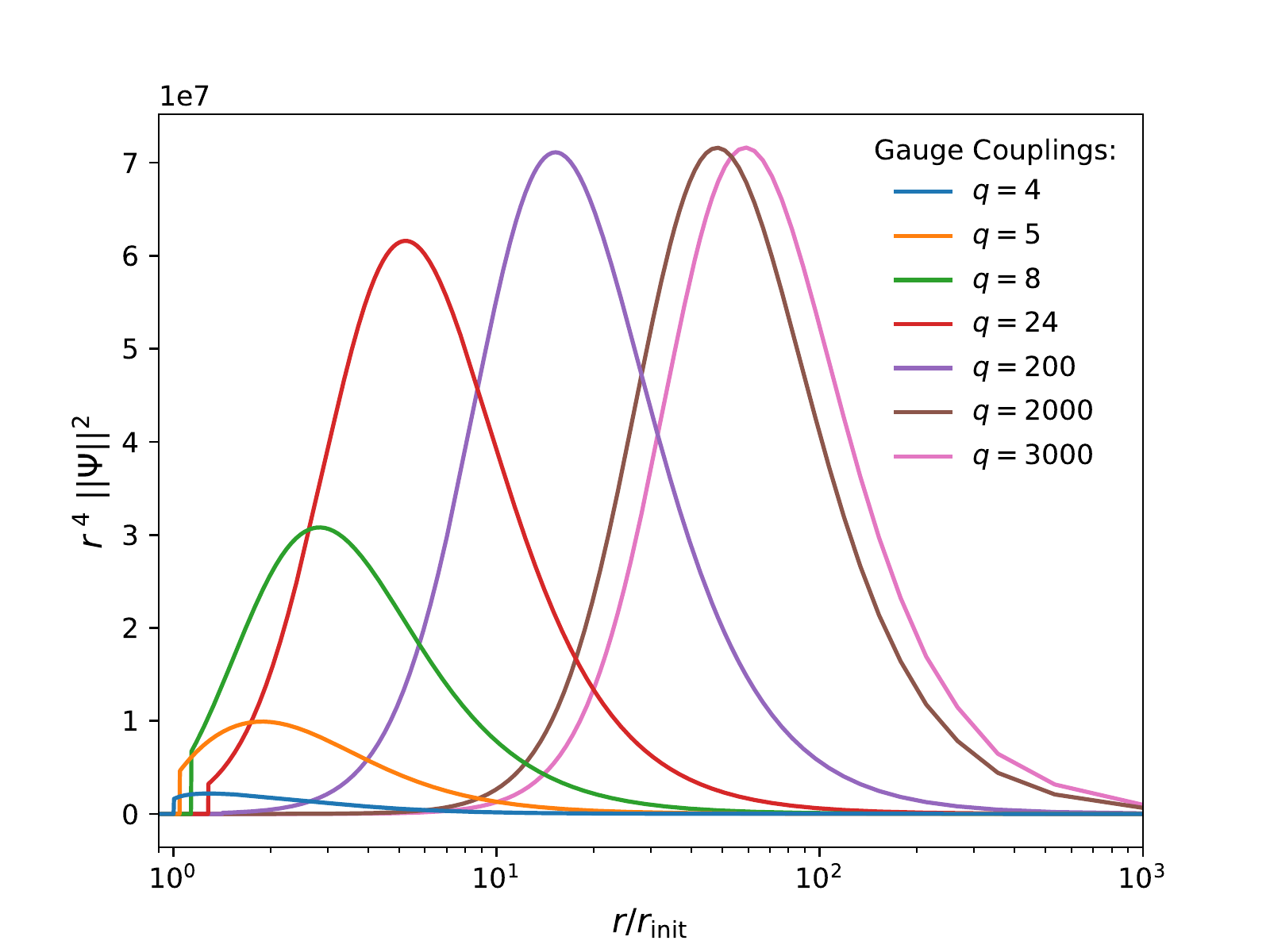}
  \caption{End state radial profile of the scalar field, for initial
    black hole with $r_+ = 100$ and $Q=0.9 Q_{\mathrm{ext}}$. For
    smaller gauge coupling $q$ the field has support closer to the
    horizon, whereas for larger $q$ the support is away from the black
    hole.\label{fig:end_states_LSR}}
\end{figure}

We experimented with varying the gauge coupling $q$;
figure~\ref{fig:AHarea_SR} shows the area of the apparent horizon as a
function of time for several different values. For larger $q$, the
final area is larger, and the growth in area happens over a much
shorter time scale. Indeed, for the smallest value, $q=4$, the area
grows by just a few percent, whereas for larger values, $q>200$, it
more than doubles. Figures~\ref{fig:charge_LSR} and
\ref{fig:end_states_LSR} show the extraction of charge and the final
radial profile of $\psi$, respectively. Indeed, the field has support closer 
to the black hole for the smaller
values of $q$, consistent with the near-horizon
instability \cite{Murata:2010dx}. For larger values of $q$, more
charge is extracted, and the field has support further from the black
hole. For very large $q$, nearly all the charge is extracted, and the
final state is nearly Schwarzschild, with a scalar condensate far
away. In all cases, the field profile has a single peak, so the
condensate is in its ground state.

\begin{figure}[tb]
  \includegraphics[width=\linewidth]{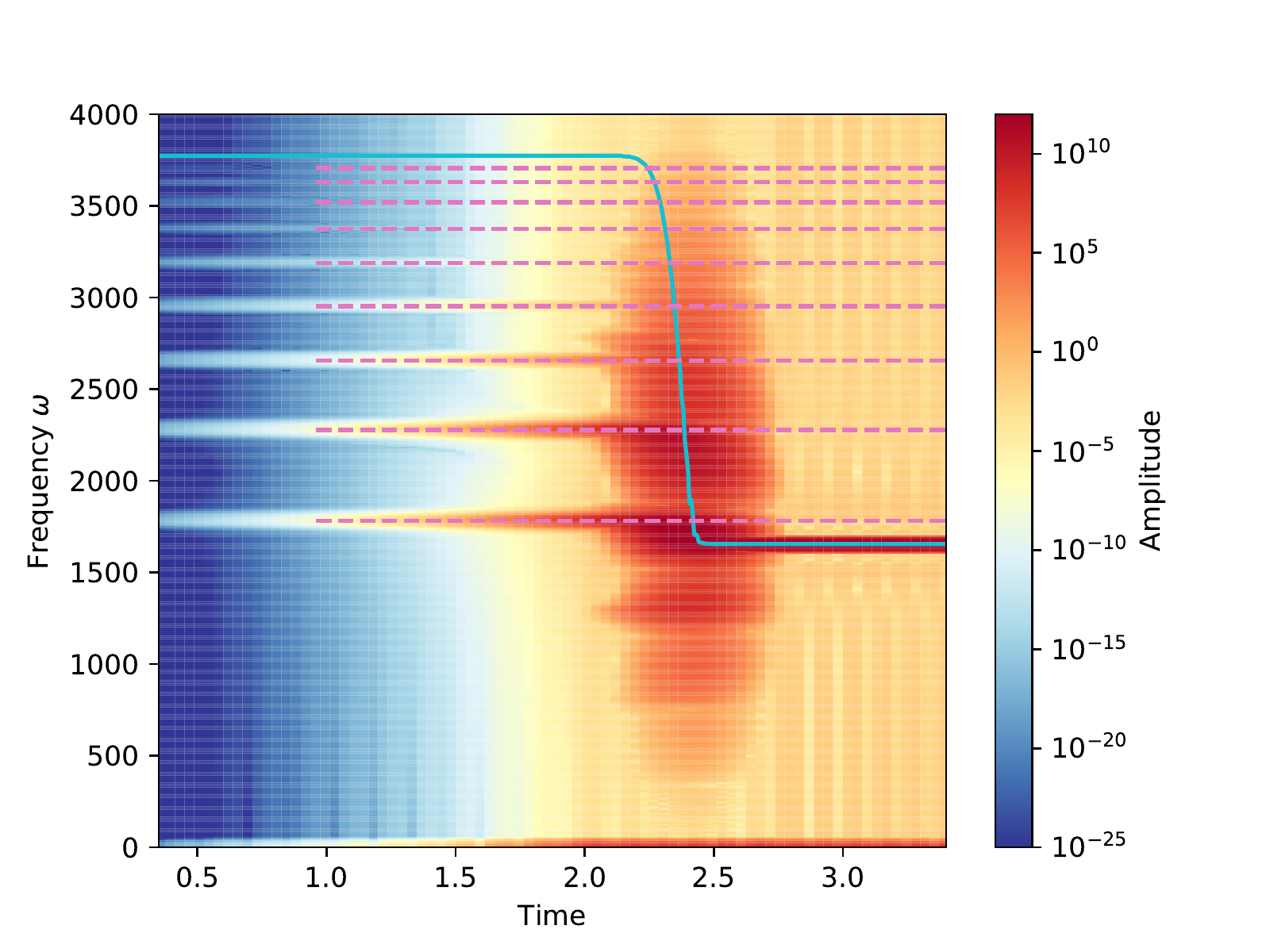}
  \caption{Spectrogram of $\Re\varphi_3(v)$ for evolution starting
    from $r_+ = 100$, $Q=0.9Q_{\text{ext}}$, and $q=24$. Initially
    there are nine unstable modes; in the end, the final state is in
    the fundamental mode. Dashed lines correspond to the quasinormal
    frequencies of table~\ref{tab:QNM_largeBH}, solid to the
    superradiant bound frequency, $-q\Phi_{\mathrm{AH}}(v)$.}
\label{fig:spectrogram_LSR}
\end{figure}

It is useful to examine also the dynamics of the boundary values of
the scalar field, $\varphi_3(v)$. We present a time-frequency analysis
in figure~\ref{fig:spectrogram_LSR} for the $q=24$ case. The peaks
correspond to quasinormal modes, and we see that at early times, there
are nine unstable modes, with the fastest growth rate for the lowest
frequency. As mass and charge are extracted, however, the superradiant
bound frequency decreases, and the higher frequency modes begin to
decay (cf.~figure~\ref{fig:merging-RN-AdS}). Eventually only the
fundamental $n=0$ mode remains. The final state is reached when the
superradiant bound frequency matches the $n=0$ mode frequency, so that
this mode becomes marginally stable. Notice the shift in the $n=0$
mode frequency over a very short time period just before saturation;
this occurs because the background solution evolves very rapidly just
before saturation, as seen in figures~\ref{fig:AHarea_SR} and
\ref{fig:charge_LSR}.

\begin{table}
{\renewcommand{\arraystretch}{1.4}
\begin{tabular}{c | c}
$n$ & $\omega$\\ \hline
 0 & $1780.01 + 17.29i$ \\
1  & $2277.33 + 16.59i$\\
2 & $2655.28 + 14.29i$\\
3 & $2952.42 + 11.97i$\\
4 & $3188.37 + 9.83i$\\
5 & $3374.87 + 7.89i$\\
6 & $3519.76 + 6.15i$\\
7 & $3628.68 + 4.55i$\\
8 & $3705.19 + 3.00i$\\
\end{tabular}
}
\caption{Quasinormal frequencies for $r_+=100$,
  $Q=0.9Q_{\mathrm{ext}}$, and $q=24$. The superradiant bound
  frequency is $-q\Phi_{\mathrm{H}}=3741.29$. The growth rate
  $\Im \omega$ decreases as the overtone number $n$
  grows.  \label{tab:QNM_largeBH} }
\end{table}

\subsection{Excited hairy black hole}\label{sec:excited}

\subsubsection{Initial data}

We have seen in the previous subsection that the final state for
generic initial data always corresponds to the fundamental
superradiant mode, even in the case where multiple unstable modes are
present. In all cases examined, the growth rate of individual modes
decreases with increasing overtone number $n$; the fundamental mode
grows fastest, as seen in table~\ref{tab:QNM_largeBH}. For
\emph{generic} initial data---with many modes initially excited---the
evolution, after possibly complicated dynamics, always comes to be
dominated by the fundamental mode.

Nevertheless, for \emph{special} initial data---with overtone modes
excited to higher amplitude---the evolution could be dominated (at
least for some time) by $n>0$ modes. If this time is longer than the
saturation time for the overtone instability, then the system will
reach the excited hairy black hole equilibrium.

To obtain suitable initial data, we require precise overtone mode
functions. To obtain these, we first select parameters $r_+$, $Q$ and
$q$ such that the background RNAdS solution has multiple unstable
modes, and then we use the method of
section~\ref{sec:linear-formalism} to calculate precise quasinormal
frequencies. We then insert the mode ansatz
$\psi(v,r)=e^{-i \omega v} R(r)$ in ingoing Eddington-Finkelstein
coordinates into~\eqref{eq:ScalarField} to obtain the radial equation
for frequency $\omega$. For each desired overtone, we integrate this
ordinary differential equation numerically, with reflecting boundary
conditions at the AdS boundary. The mode can then be taken as initial
data at advanced time $v=0$.
 
\begin{figure}[tb]
  \includegraphics[width=\linewidth]{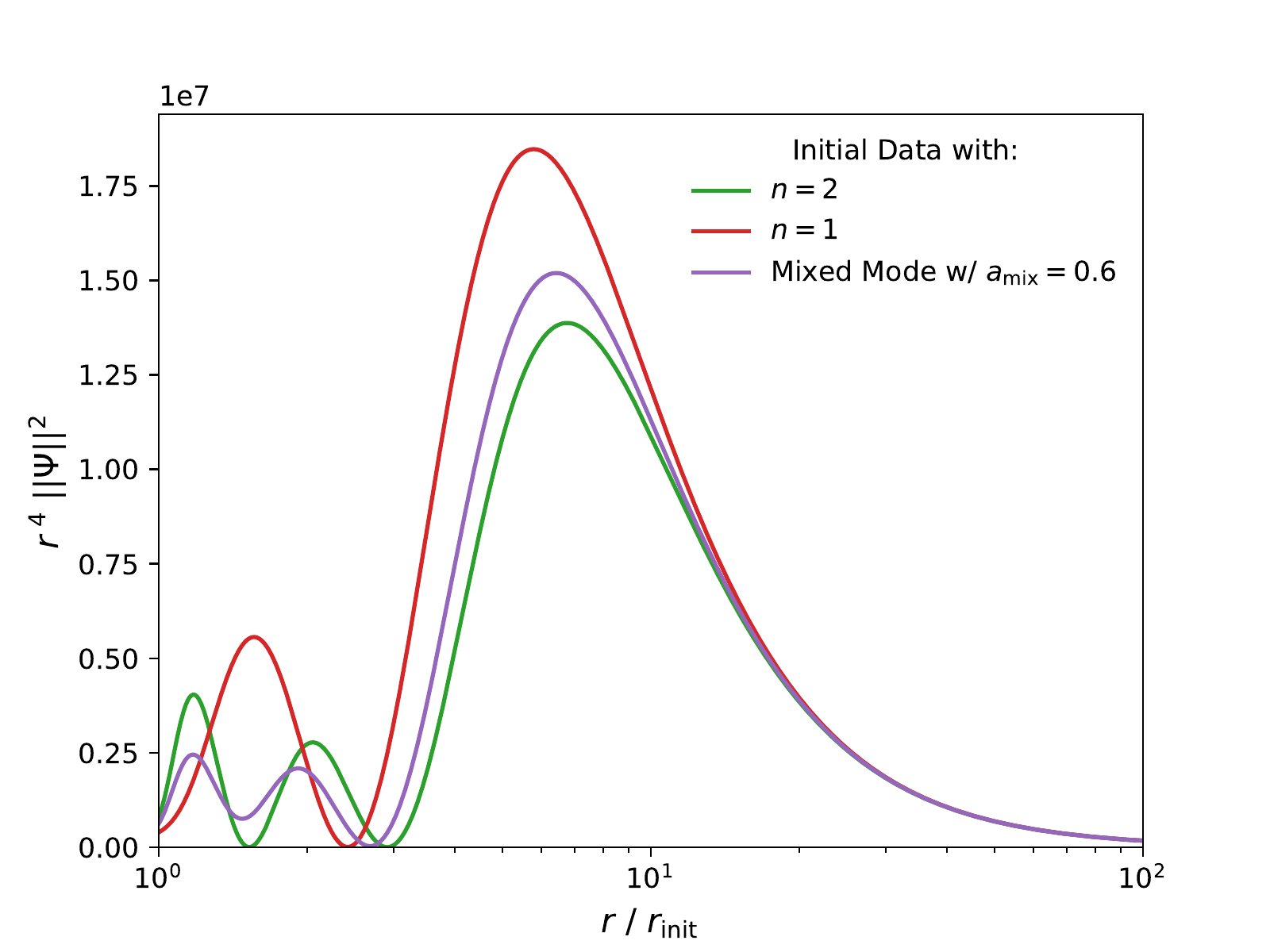}
  \caption{Radial profiles of initial data for initial values
    $r_+=100$, $Q=0.9 Q_{\mathrm{ext}}$, and $q=11$. Two curves
    correspond to single-mode data, and the third to mixed-mode
    data.\label{fig:InitialData}}
\end{figure}

We consider two types of special initial data. The first consists of a
single overtone mode $\psi_n$, which, if pure enough, we expect to
evolve into an excited hairy black hole. The second type of initial
data is a mixture of two modes, i.e.,
\begin{equation}\label{eq:mixeddata}
  \psi_{\text{mix}} = a_{\text{mix}} \psi_2 + (1-a_{\text{mix}}) \psi_1 ,
\end{equation}
with, e.g., $a_{\text{mix}} = 0.999$, and the amplitudes normalized using
the infinity norm. 
With these data, we hope to achieve a cascade, where
initially a $n=2$ excited black hole forms, which then decays to
$n=1$, and then $n=0$. Some initial data profiles are shown in
figure~\ref{fig:InitialData}.

\subsubsection{Results}

As in the generic evolution, we fix $r_+=100$, $Q=0.9Q_{\text{ext}}$,
and $L=1$. We then consider two cases for the scalar field charge,
$q=8, 11$. The significance of these latter two choices is that for
$q=8$, the background RNAdS solution has two unstable modes, whereas
for $q=11$, there are three unstable modes.

\textbf{\underline{q=8:}} In this case, modes $n=0,1$ are unstable,
with initial quasinormal frequencies $\omega_0= 978.70 + 7.34i$ and
$\omega_1= 1168.69 + 4.79i$. To obtain the $n=1$ excited hairy black
hole, we take initial data to consist of $\psi_1$. We find that under
evolution, the mode grows exponentially and extracts charge and mass
from the black hole, similar to the generic evolution. This causes the
superradiant bound frequency, $-q\Phi_{\text{H}}$, to drop until it
matches $\Re \omega_1$. (The mode frequency $\Re \omega_1$ evolves due
to the changing background spacetime, but this is negligible compared
to the change in superradiant bound frequency.)  At this point,
superradiance stops, and the system settles into the excited hairy
black hole state. The black hole is static, with the scalar field
oscillating harmonically.

\begin{figure}[tbh]
  \includegraphics[width=\linewidth]{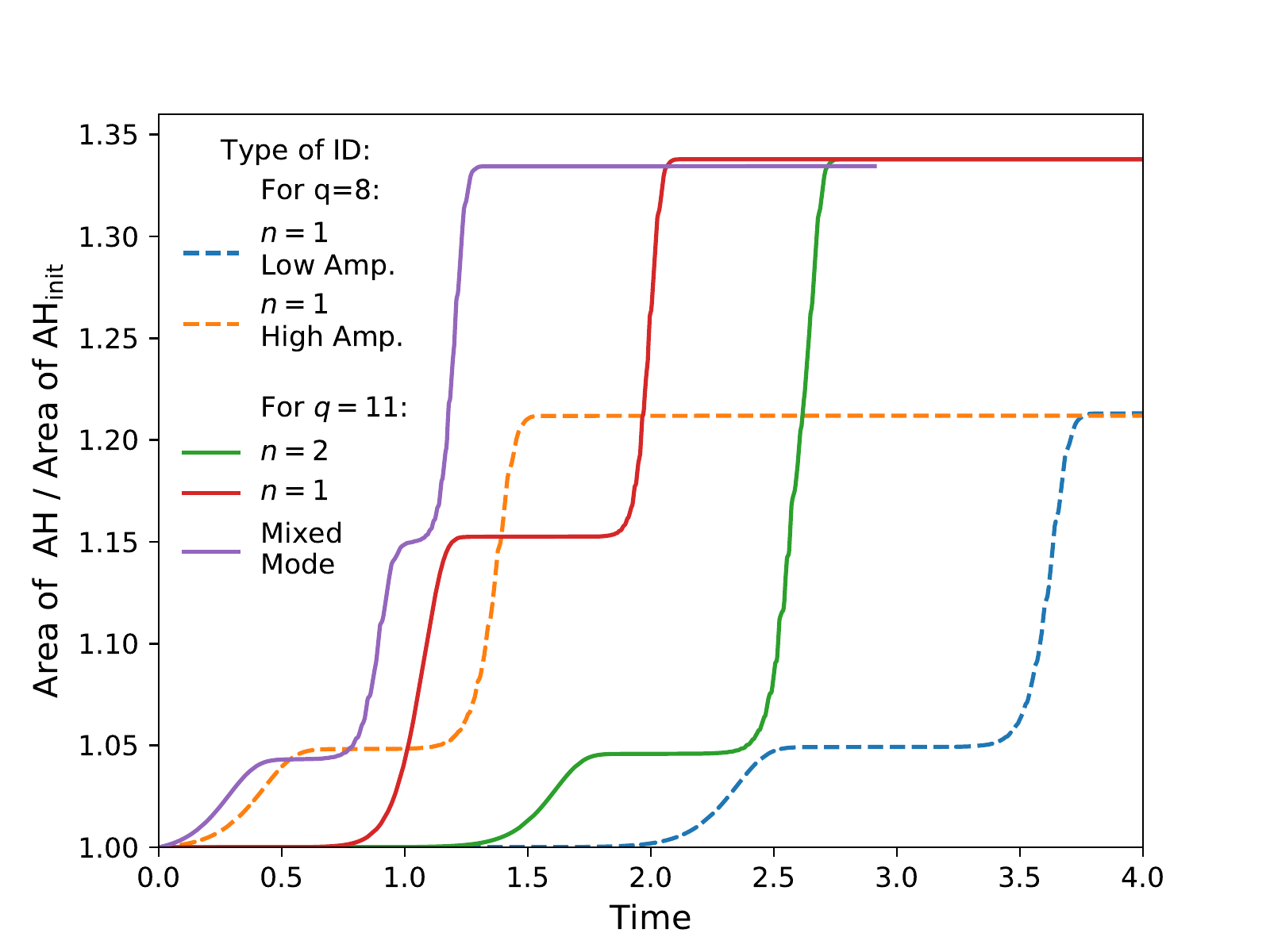}
  \caption{Normalized area of the apparent horizon for initial
    $r_+=100$ and $Q=0.9Q_{\text{ext}}$. Dashed curves correspond to
    $q=8$, solid to $q=11$. Excited hairy black hole solutions occur
    at the temporary plateaus.\label{fig:AHarea}}
\end{figure}
The area of the apparent horizon is shown in figure~\ref{fig:AHarea}
(either one of the dashed curves). The excited hairy black hole is
seen as a plateau, where the area stops growing because the scalar
field is no longer extracting mass and charge. To this point, the
description parallels that of section~\ref{sec:generic}. However,
after some time, the area begins to grow \emph{again}; this is because
the $n=0$ mode was present and growing the entire time. Indeed, since
$\Re\omega_0 < \Re\omega_1$, the fundamental mode remains
superradiantly unstable even after the overtone saturates. When the
amplitude of the $n=0$ mode becomes large, it disrupts the static
black hole and causes its area to grow significantly. Once
$-q\Phi_{\text{H}}$ drops below $\Re\omega_1$, the overtone mode falls
back into the black hole, and once it reaches $\Re \omega_0$,
superradiance stops completely. At this point, the black hole is in
its final state, described by the $n=0$ ground state mode.

It is impossible to avoid triggering the $n=0$ mode. At the initial
time, the data for $\psi_1$ will always have numerical error, which
will have some overlap with $\psi_0$. Moreover, during evolution,
$\psi_0$ will be excited nonlinearly. To determine the origin of the
observed $n=0$ mode, we varied the initial perturbation amplitude,
and read off the times $t_1$ and $t_0$ at which the $n=1$ mode
saturates and the $n=0$ mode overtakes the dynamics,
respectively. Using the growth rates from the linear analysis we know
that
\begin{equation}
  \log \left(\frac{A_1}{A_0}\right) = t_0 \Im{\omega_0} - t_1 \Im{\omega_1} \, ,
\label{eq:seeding_n0}
\end{equation}
where $A_1$ and $A_0$ are the initial amplitudes respectively.
We used this formula
to calculate the amplitude $A_0$, given $A_1$ and the measured $t_0$,
$t_1$.

For sufficiently small $A_1$, the calculated $A_0$ has only a mild
dependence on $A_1$ indicating the zero mode is sourced primarily by
truncation. (This was confirmed by noting the onset of this behavior
depends on the resolution, with finer resolutions showing such
behavior at smaller values of $A_1$.) However, for
$A_1 \gtrsim 5 \times10^{-3}$, we found that
\begin{equation}\label{eq:scalingA}
  A_0 \sim A_1^{\phantom{1}2.75\pm 0.07},
\end{equation}
This value is consistent with the seed arising from the
self-gravitating contribution of the scalar field
($A_0 \propto A_1^{\phantom{1}3}$).

Evolutions with ``low'' and ``high'' initial perturbation amplitudes
are depicted in figure~\ref{fig:AHarea}.  Notice that although the
saturation times differ between the two cases, the areas of the hairy
black holes are largely independent of the amplitude of the initial
data, as long as the amplitude is low.

\textbf{\underline{q=11:}} For $q=11$, modes $n=0,1,2$ are unstable,
with frequencies $\omega_0=1174.13 + 9.94i$,
$\omega_1 = 1448.56+ 8.06i $, and $\omega_2 = 1615.34 + 5.20i$. We
therefore consider three types of initial data, data with modes $n=1$
and $n=2$ individually excited, and the mixed-mode initial
data~\eqref{eq:mixeddata}. Simulation results for the apparent horizon
area are shown as solid curves in figure~\ref{fig:AHarea}.

\begin{figure}[tbh]
  \includegraphics[width=\linewidth]{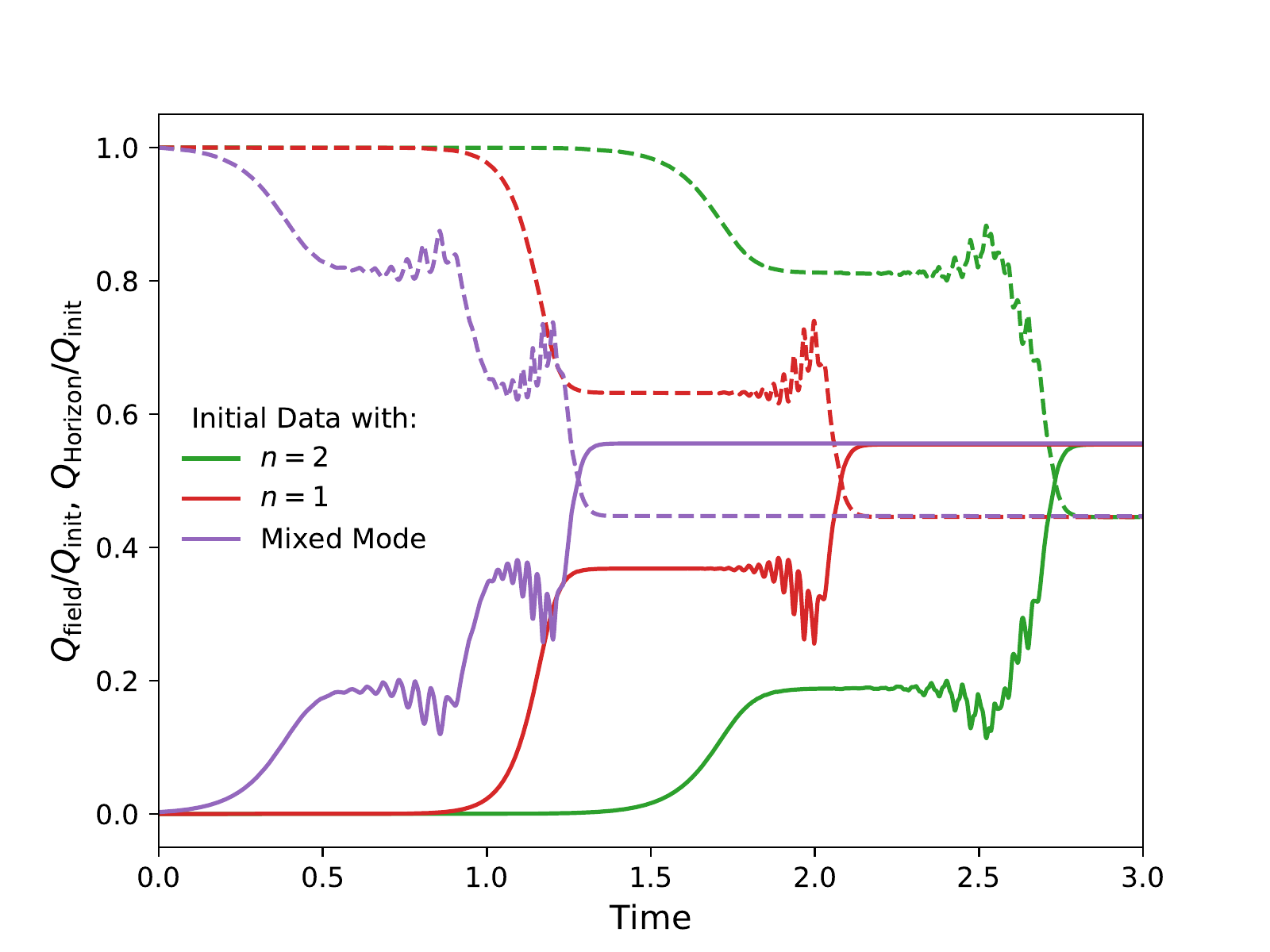}
  \caption{Charge transfer as a function of time for initial data with
    $r_+ = 100$, $Q=0.9Q_{\text{ext}}$, and $q=11$. Solid curves
    are the (normalized) charge of the scalar field, dashed curves the
    charge of the black hole. 
    \label{fig:charge_transfer}}
\end{figure}

The behavior for single-mode initial data is qualitatively similar to
$q=8$. We find, however, that the area of the $n=1$ excited hairy
black hole is larger than the $n=2$ black hole, consistent with the
discussion above and $\Re \omega_1 < \Re \omega_2$. The final black
hole is the same in both single-mode cases. In
figure~\ref{fig:charge_transfer} we plot the electric charge of the
black hole and the scalar field. This shows that at the end of the
excited hairy black hole life, significant amounts of charge are
deposited back into the hole. This corresponds to the rapid decay of
overtone hair as the superradiant bound frequency drops below the
overtone frequency (cf.~figure~\ref{fig:largebhspec}, where
quasinormal mode decay time scales are much shorter than growth time
scales).

For mixed-mode initial data, we take mixing ratio
$a_{\mathrm{mix}}=0.999$, i.e., the data are $99.9\%$ $n=2$ and
$0.1\%$ $n=1$.  This allows the $n=2$ mode to dominate the dynamics
for early times. The length of time the $n=2$ will dominate can be
estimated, using a similar calculation to~\eqref{eq:seeding_n0}, to be
$\Delta t = (\log(a_{\mathrm{mix}}/(1-a_{\mathrm{mix}})) +
t_{\mathrm{s}}(\Im{\omega_2}-\Im{\omega_1}))/\Im{\omega_1} \sim 0.66$,
where $t_{\mathrm{s}}$ is the time at which the $n=2$ mode saturates.
Indeed, we observe (purple solid curve in figures~\ref{fig:AHarea} and
\ref{fig:charge_transfer}) that the system cascades through \emph{two}
transient excited hairy black hole states, first $n=2$, then $n=1$,
before settling in the ground state. The hairy black hole states match
those seen in the single-mode evolutions.

\begin{figure}[tbh]
  \includegraphics[width=\linewidth]{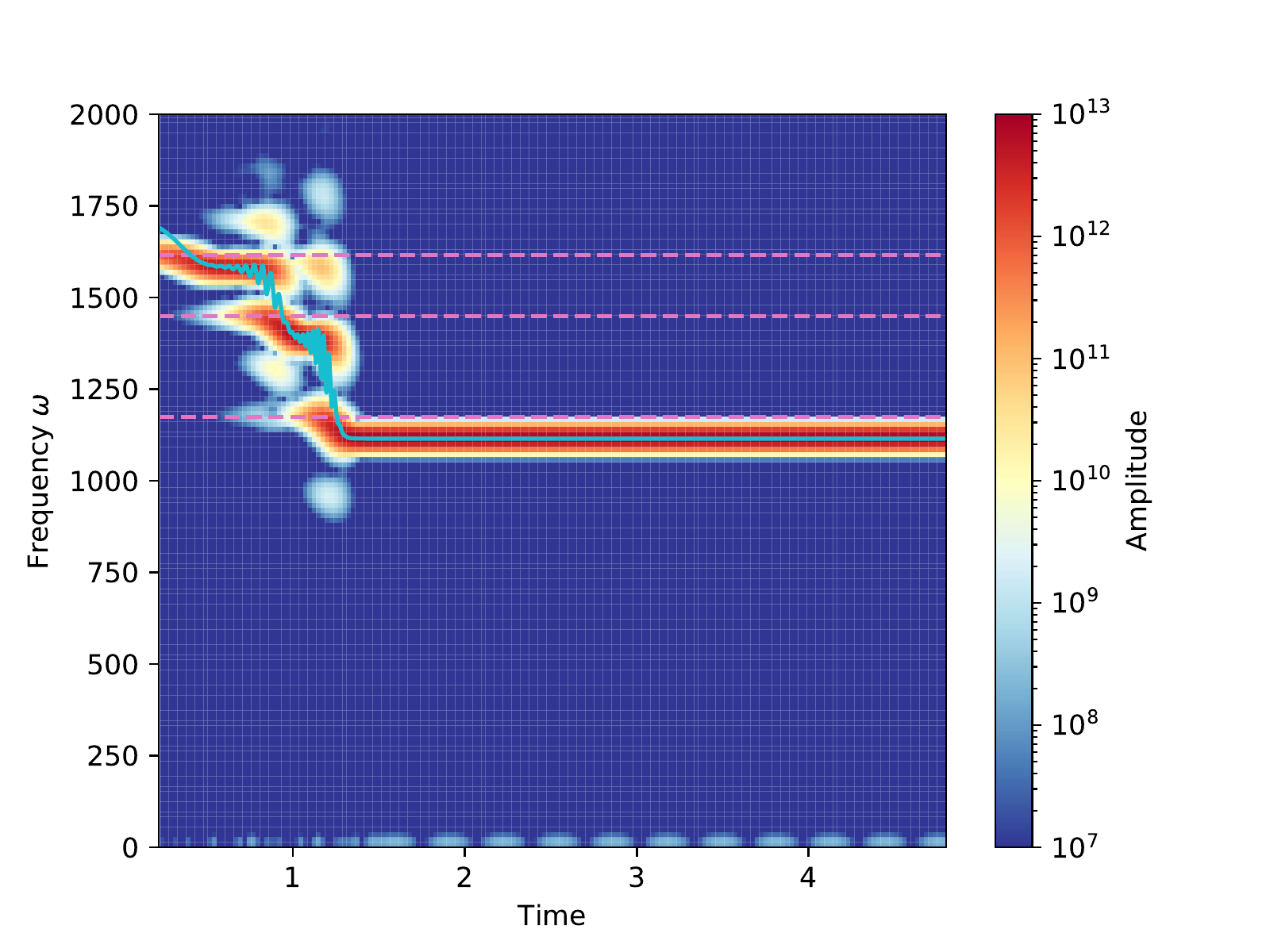}
  \caption{Spectrogram of $\varphi_3(v)$ for mixed mode initial data,
    with $r_+=100$, $Q=0.9Q_{\text{ext}}$, $q=11$, and
    $a_{\mathrm{Mix}}=0.999$. This shows a cascade through hairy black
    holes with $n=2 \rightarrow 1 \rightarrow 0$. Dashed lines indicate 
    values of the initial mode frequencies (calculated
    using the linear analysis), solid corresponds to the superradiant bound
$-q\Phi_{\mathrm{AH}}(v)$. \label{fig:double_transition}}
\end{figure}

We present a spectrogram for the mixed-mode evolution in
figure~\ref{fig:double_transition}. This shows a clear progression
through the three unstable modes. Notice again that the final $n=0$
oscillation frequency is slightly lower than the frequency of the
initial $n=0$ quasinormal mode. This shift arises because the final
black hole is different from the initial one, and the superradiant
bound frequency has shifted.

\section{Conclusions}\label{sec:conclusion}

In this work we computed the $l=0$ charged scalar quasinormal mode
spectrum for RNAdS, and in cases of unstable modes, we numerically
simulated the full nonlinear development.

The quasinormal mode analysis used the continued fraction method,
which enabled us to study regions of parameter space that were not
previously examined due to a lack of small parameter needed for
analytic studies. We showed in particular that for large black holes,
the zero-damped mode family can become superradiantly unstable, and
exhibits behavior similar to the small black hole AdS mode
family. Furthermore, the leading unstable mode is identified with the
near-horizon condensation instability.

At the nonlinear level, we studied the evolution of these large-RNAdS
unstable modes. We showed that the generic end point is a static black
hole with a (harmonically-oscillating) scalar condensate, similar to
earlier results for small~\cite{Bosch:2016vcp} and
planar~\cite{Murata:2010dx} RNAdS. We also showed that for black holes
with multiple unstable modes, special initial data can be chosen that
evolve to a transient excited hairy black hole solution before
decaying to the generic end state.

It is tempting to draw an analogy between classical hairy black hole
energy levels and quantum energy levels of atoms. In this picture (in
AdS) the scalar field can only exchange energy (and charge and angular
momentum) with the black hole, so the horizon plays the role of the
atomic environment. In the black hole case, however, level transitions
can only occur in the direction of decreasing overtone
number. Transitions in the reverse direction are forbidden by the area
theorem.

The reason that the final hairy black hole is always in the $n=0$
configuration is because out of all quasinormal modes, the $n=0$ mode has
lowest $\Re\omega > 0$. The superradiance condition is
$0 < \Re \omega < -q\Phi_{\text{H}}$, and as mass and charge
extraction cause the upper bound to decrease, the $n=0$ mode is the
last to remain unstable. We were nevertheless able to obtain the
transient excited hairy black holes because the instability growth
rates of the overtones are comparable and we were free to choose
special overtone initial data.

Had the growth rate of overtone modes been \emph{higher} than the
fundamental mode, the situation would be somewhat different. Although
the final configuration would be unchanged (because of the ordering of
the real parts of the frequencies), the excited hairy black hole
states would occur transiently for \emph{generic} initial data. This
reverse ordering of overtone growth rates occurs for superradiantly-unstable
angular harmonics of Proca fields in Kerr~\cite{Siemonsen:2019ebd},
which is relevant to searches for ultralight dark
matter~\cite{Arvanitaki:2014wva}. It would be interesting to study any
observational consequences of transient overtone equilibria in this
context. 

Another context where the interplay between instability criteria and
growth rates leads to transient states in generic evolutions is the
superradiant instability of Kerr-AdS. These states, however, involve
different angular harmonics rather than radial overtones. Indeed
recent simulations~\cite{Chesler:2018txn} of the Kerr-AdS superradiant
instability show an evolution dominated by a series of epochs
consisting of black resonators~\cite{Dias:2015rxy}, which are
themselves unstable~\cite{Green:2015kur}.

Instability of RNAdS and subsequent hairy black hole formation has
been proposed as a holographic dual to a superconducting phase
transition~\cite{Hartnoll:2008kx,Horowitz:2011dz}. It is intriguing
to seek also a holographic interpretation of the transient hairy black
hole equilibria that we uncovered.

More generally, our work underscores the importance of overtone modes
and nonlinear effects in black hole perturbations. For perturbed black
holes arising from a binary merger, recent
works~\cite{Giesler:2019uxc,Isi:2019aib,Ota:2019bzl} have argued that
the post-merger gravitational-wave signal can be well-described by a
combination of overtone modes evolving linearly in a Kerr
background. Other works, however, have argued for additional nonlinear
mode excitation~\cite{Zlochower:2003yh,East:2013mfa}, sometimes
through parametric instabilities~\cite{Yang:2014tla,Yang:2015jja}. For
weakly perturbed black holes, we measured a natural scaling
\eqref{eq:scalingA} that describes nonlinear mode excitation in
RNAdS. Further work and numerical simulations will be needed to build
intuition and understand the validity of linear analyses in strongly
perturbed regimes.

\begin{acknowledgments}

We would like to thank P. Zimmerman and W. East for discussions and comments
throughout this project. This work was supported in part by CONACyT-Mexico (P.B.), NSERC,
and CIFAR (L.L.).  H.R. thanks the Perimeter Institute for Theoretical Physics
for hospitality and accommodations during an internship sponsored by the \'Ecole
Normale Sup\'erieure. This research was supported in part by Perimeter Institute
for Theoretical Physics. Research at Perimeter Institute is supported by the
Government of Canada and by the Province of Ontario through the Ministry of
Research, Innovation and Science.

\end{acknowledgments}

\appendix

\section{Code validation}

\begin{figure}[htb]
	\includegraphics[width=0.5\textwidth]{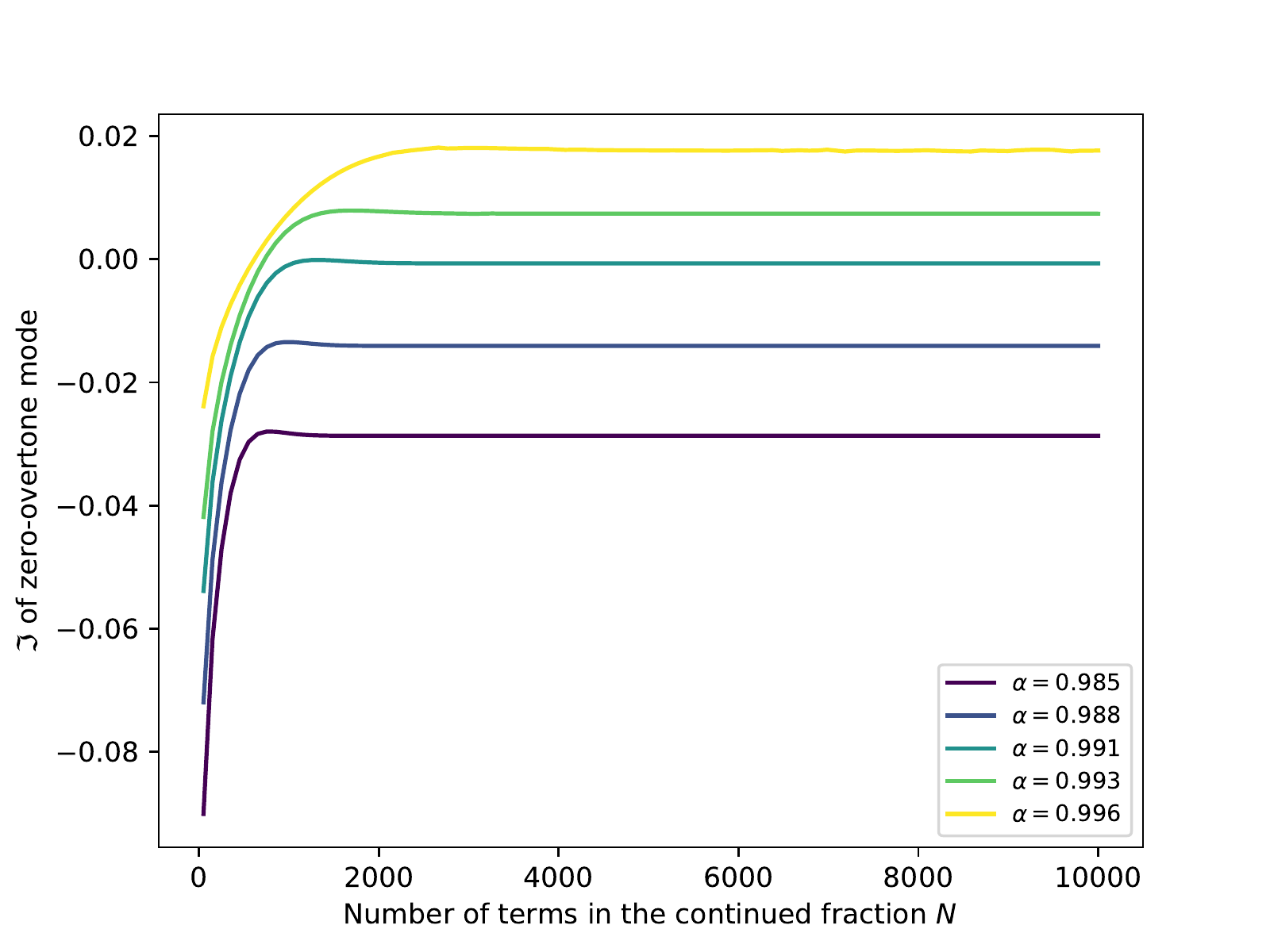}
	\caption{Imaginary part of the NH mode as a function of the number of terms used in the continued fraction $N$, for different values of the extremality parameter $a$. We can observe that convergence is more difficult when $a$ gets closer to 1, as the equations used break down.}
\label{fig:ContinuedFractionConvergence}
\end{figure}

We have confirmed the validity of both codes used extensively in
this work through self-convergence tests as well as comparison with
available results in suitable regimes. With regards to self-convergence tests,
we have verified that as the number of terms employed in the continued fraction
method is increased, our results asymptote to consistent results and, that typically
this takes place when $N \gtrapprox 3000$ (see figure~\ref{fig:ContinuedFractionConvergence}). 
Convergence of the non-linear code has been recently demonstrated in~\cite{Bosch:2016vcp}.
As mentioned, we also compare with specific results; in particular: our QNM frequencies obtained in our 
linear analysis agree
with those obtained in~\cite{Horowitz:1999jd} in the Schwarzschild-AdS limit
($Q=0$,$q=0$) to better than $1.2\%$ for the real part of $\omega$ and better
than $0.5\%$ for the imaginary part of $\omega$. In the charged, small black hole case, 
our results agree with those presented in~\cite{Uchikata:2011zz} to better than  $0.1\%$. An
example of these comparisons is given in Tables~\ref{qnmtable1},\ref{qnmtable2}.
In the large black hole, small charge regime, our results agree with those in~\cite{Berti:2003ud}. 
to better than $7.5\%$ and $4\%$ for the real and imaginary parts of $\omega$ respectively.
We have also confirmed solutions obtained with our full non-linear simulations illustrate
initial growth rates --in the unstable regime-- and black hole QNMs consistent with the expected
results from our linear studies. 

\begin{table}
\begin{tabular}{|l|l|l|l|l|}
	\hline
	& \multicolumn{2}{c|}{Values of \cite{Horowitz:1999jd}} & \multicolumn{2}{c|}{Our values}\\
	\hline
	$r_+$ & $\Re(\omega)$ & $\Im(\omega)$ & $\Re(\omega)$ & $\Im(\omega)$\\
	\hline
	0.4 & \num{2.3629} & \num{-1.0064} & \num{2.3629} & \num{-1.0065} \\
	\hline
	0.6 & \num{2.4316} & \num{-1.5797} & \num{2.4316} & \num{-1.5797} \\
	\hline
	0.8 & \num{2.5878} & \num{-2.1304} & \num{2.5878} & \num{-2.1304} \\
	\hline
	1 & \num{2.7982} & \num{-2.6712} & \num{2.7982} & \num{-2.6712} \\
	\hline
	5 & \num{9.4711} & \num{-13.3255} & \num{9.4711} & \num{-13.3255} \\
	\hline
	10 & \num{18.6070} & \num{-26.6418} & \num{18.6070} & \num{-26.6418} \\
	\hline
	50 & \num{92.4937} & \num{-133.1933} & \num{92.4937} & \num{-133.1933} \\
	\hline
	100 & \num{184.9534} & \num{-266.3856} & \num{184.9534} & \num{-266.3856} \\
	\hline
\end{tabular}
\caption{QNMs found by the linear analysis for a Schwarzschild black hole ($a = 0$, $q=0$) and comparison with the results of \cite{Horowitz:1999jd} (Table 1). We can see our values agree completely with the previous results.}
\label{qnmtable1}
\end{table}

\begin{table}
\begin{tabular}{|l|l|l|l|l|l|}
	\hline
	& & \multicolumn{2}{c|}{Values of \cite{Uchikata:2011zz}} & \multicolumn{2}{c|}{Our values}\\
	\hline
	$q$ & $a$ & $\Re(\omega)$ & $\Im(\omega)$ & $\Re(\omega)$ & $\Im(\omega)$\\
	\hline
	0 & 0 & \num{2.6928} & \num{-1.0095e-1} & \num{2.6928} & \num{-1.0096e-1} \\
	\cline{2-6} & 0.2 & \num{2.6801} & \num{-1.0434e-1} & \num{2.6801} & \num{-1.0434e-1} \\
	\cline{2-6} & 0.4 & \num{2.6410} & \num{-1.1625e-1} & \num{2.6411} & \num{-1.1626e-1} \\
	\cline{2-6} & 0.6 & \num{2.5723} & \num{-1.4417e-1} & \num{2.5723} & \num{-1.4417e-1} \\
	\cline{2-6} & 0.8 & \num{2.4787} & \num{-2.0756e-1} & \num{2.4788} & \num{-2.0757e-1} \\
	\cline{2-6} & 0.9 & \num{2.4332} & \num{-2.5493e-1} & \num{2.4332} & \num{-2.5493e-1} \\
	\hline
	2 & 0.2 & \num{2.7614} & \num{-8.1233e-2} & \num{2.7615} & \num{-8.1233e-2} \\
	\cline{2-6} & 0.4 & \num{2.8058} & \num{-6.7765e-2} & \num{2.8059} & \num{-6.7765e-2} \\
	\cline{2-6} & 0.6 & \num{2.8256} & \num{-5.8950e-2} & \num{2.8257} & \num{-5.8950e-2} \\
	\cline{2-6} & 0.8 & \num{2.8161} & \num{-5.6894e-2} & \num{2.8161} & \num{-5.6894e-2} \\
	\cline{2-6} & 0.9 & \num{2.7977} & \num{-6.3772e-2} & \num{2.7978} & \num{-6.3773e-2} \\
	\hline
	4 & 0.2 & \num{2.8414} & \num{-6.1698e-2} & \num{2.8415} & \num{-6.1698e-2} \\
	\cline{2-6} & 0.4 & \num{2.9650} & \num{-3.4408e-2} & \num{2.9650} & \num{-3.4408e-2} \\
	\cline{2-6} & 0.6 & \num{3.0672} & \num{-1.3945e-2} & \num{3.0672} & \num{-1.3946e-2} \\
	\cline{2-6} & 0.8 & \num{3.1515} & \num{1.7314e-3} & \num{3.1515} & \num{1.7314e-3} \\
	\cline{2-6} & 0.9 & \num{3.1878} & \num{6.5742e-3} & \num{3.1879} & \num{6.5744e-3} \\
	\hline
\end{tabular}
\caption{QNMs found by the linear analysis for a small black hole ($r_+ = 0.1$) and comparison with the results of \cite{Uchikata:2011zz} (Table 2). We can see our values agree completely with the previous results.}
\label{qnmtable2}
\end{table}
\clearpage
\bibliography{main}

\begin{thebibliography}{51}%
\makeatletter
\providecommand \@ifxundefined [1]{%
 \@ifx{#1\undefined}
}%
\providecommand \@ifnum [1]{%
 \ifnum #1\expandafter \@firstoftwo
 \else \expandafter \@secondoftwo
 \fi
}%
\providecommand \@ifx [1]{%
 \ifx #1\expandafter \@firstoftwo
 \else \expandafter \@secondoftwo
 \fi
}%
\providecommand \natexlab [1]{#1}%
\providecommand \enquote  [1]{``#1''}%
\providecommand \bibnamefont  [1]{#1}%
\providecommand \bibfnamefont [1]{#1}%
\providecommand \citenamefont [1]{#1}%
\providecommand \href@noop [0]{\@secondoftwo}%
\providecommand \href [0]{\begingroup \@sanitize@url \@href}%
\providecommand \@href[1]{\@@startlink{#1}\@@href}%
\providecommand \@@href[1]{\endgroup#1\@@endlink}%
\providecommand \@sanitize@url [0]{\catcode `\\12\catcode `\$12\catcode
  `\&12\catcode `\#12\catcode `\^12\catcode `\_12\catcode `\%12\relax}%
\providecommand \@@startlink[1]{}%
\providecommand \@@endlink[0]{}%
\providecommand \url  [0]{\begingroup\@sanitize@url \@url }%
\providecommand \@url [1]{\endgroup\@href {#1}{\urlprefix }}%
\providecommand \urlprefix  [0]{URL }%
\providecommand \Eprint [0]{\href }%
\providecommand \doibase [0]{http://dx.doi.org/}%
\providecommand \selectlanguage [0]{\@gobble}%
\providecommand \bibinfo  [0]{\@secondoftwo}%
\providecommand \bibfield  [0]{\@secondoftwo}%
\providecommand \translation [1]{[#1]}%
\providecommand \BibitemOpen [0]{}%
\providecommand \bibitemStop [0]{}%
\providecommand \bibitemNoStop [0]{.\EOS\space}%
\providecommand \EOS [0]{\spacefactor3000\relax}%
\providecommand \BibitemShut  [1]{\csname bibitem#1\endcsname}%
\let\auto@bib@innerbib\@empty
%</preamble>
\bibitem [{\citenamefont {Choptuik}(1993)}]{Choptuik:1992jv}%
  \BibitemOpen
  \bibfield  {author} {\bibinfo {author} {\bibfnamefont {Matthew~W.}\
  \bibnamefont {Choptuik}},\ }\bibfield  {title} {\enquote {\bibinfo {title}
  {{Universality and scaling in gravitational collapse of a massless scalar
  field}},}\ }\href {\doibase 10.1103/PhysRevLett.70.9} {\bibfield  {journal}
  {\bibinfo  {journal} {Phys. Rev. Lett.}\ }\textbf {\bibinfo {volume} {70}},\
  \bibinfo {pages} {9--12} (\bibinfo {year} {1993})}\BibitemShut {NoStop}%
%%CITATION = PRLTA,70,9;%%
\bibitem [{\citenamefont {Gundlach}(1999)}]{Gundlach:1999cu}%
  \BibitemOpen
  \bibfield  {author} {\bibinfo {author} {\bibfnamefont {Carsten}\ \bibnamefont
  {Gundlach}},\ }\bibfield  {title} {\enquote {\bibinfo {title} {{Critical
  phenomena in gravitational collapse}},}\ }\href@noop {} {\bibfield  {journal}
  {\bibinfo  {journal} {Living Rev. Rel.}\ }\textbf {\bibinfo {volume} {2}},\
  \bibinfo {pages} {4} (\bibinfo {year} {1999})},\ \Eprint
  {http://arxiv.org/abs/gr-qc/0001046} {arXiv:gr-qc/0001046 [gr-qc]}
  \BibitemShut {NoStop}%
%%CITATION = GR-QC/0001046;%%
\bibitem [{\citenamefont {Bizon}\ and\ \citenamefont
  {Rostworowski}(2011)}]{Bizon:2011gg}%
  \BibitemOpen
  \bibfield  {author} {\bibinfo {author} {\bibfnamefont {Piotr}\ \bibnamefont
  {Bizon}}\ and\ \bibinfo {author} {\bibfnamefont {Andrzej}\ \bibnamefont
  {Rostworowski}},\ }\bibfield  {title} {\enquote {\bibinfo {title} {{On weakly
  turbulent instability of anti-de Sitter space}},}\ }\href {\doibase
  10.1103/PhysRevLett.107.031102} {\bibfield  {journal} {\bibinfo  {journal}
  {Phys.~Rev.~Lett.}\ }\textbf {\bibinfo {volume} {107}},\ \bibinfo {pages}
  {031102} (\bibinfo {year} {2011})},\ \Eprint {http://arxiv.org/abs/1104.3702}
  {arXiv:1104.3702 [gr-qc]} \BibitemShut {NoStop}%
%%CITATION = ARXIV:1104.3702;%%
\bibitem [{\citenamefont {Carrasco}\ \emph {et~al.}(2012)\citenamefont
  {Carrasco}, \citenamefont {Lehner}, \citenamefont {Myers}, \citenamefont
  {Reula},\ and\ \citenamefont {Singh}}]{Carrasco:2012nf}%
  \BibitemOpen
  \bibfield  {author} {\bibinfo {author} {\bibfnamefont {Federico}\
  \bibnamefont {Carrasco}}, \bibinfo {author} {\bibfnamefont {Luis}\
  \bibnamefont {Lehner}}, \bibinfo {author} {\bibfnamefont {Robert~C.}\
  \bibnamefont {Myers}}, \bibinfo {author} {\bibfnamefont {Oscar}\ \bibnamefont
  {Reula}}, \ and\ \bibinfo {author} {\bibfnamefont {Ajay}\ \bibnamefont
  {Singh}},\ }\bibfield  {title} {\enquote {\bibinfo {title} {{Turbulent flows
  for relativistic conformal fluids in 2+1 dimensions}},}\ }\href {\doibase
  10.1103/PhysRevD.86.126006} {\bibfield  {journal} {\bibinfo  {journal}
  {Phys.~Rev.}\ }\textbf {\bibinfo {volume} {D86}},\ \bibinfo {pages} {126006}
  (\bibinfo {year} {2012})},\ \Eprint {http://arxiv.org/abs/1210.6702}
  {arXiv:1210.6702 [hep-th]} \BibitemShut {NoStop}%
%%CITATION = ARXIV:1210.6702;%%
\bibitem [{\citenamefont {Yang}\ \emph
  {et~al.}(2015{\natexlab{a}})\citenamefont {Yang}, \citenamefont {Zimmerman},\
  and\ \citenamefont {Lehner}}]{Yang:2014tla}%
  \BibitemOpen
  \bibfield  {author} {\bibinfo {author} {\bibfnamefont {Huan}\ \bibnamefont
  {Yang}}, \bibinfo {author} {\bibfnamefont {Aaron}\ \bibnamefont {Zimmerman}},
  \ and\ \bibinfo {author} {\bibfnamefont {Luis}\ \bibnamefont {Lehner}},\
  }\bibfield  {title} {\enquote {\bibinfo {title} {{Turbulent Black Holes}},}\
  }\href {\doibase 10.1103/PhysRevLett.114.081101} {\bibfield  {journal}
  {\bibinfo  {journal} {Phys. Rev. Lett.}\ }\textbf {\bibinfo {volume} {114}},\
  \bibinfo {pages} {081101} (\bibinfo {year} {2015}{\natexlab{a}})},\ \Eprint
  {http://arxiv.org/abs/1402.4859} {arXiv:1402.4859 [gr-qc]} \BibitemShut
  {NoStop}%
%%CITATION = ARXIV:1402.4859;%%
\bibitem [{\citenamefont {Adams}\ \emph {et~al.}(2014)\citenamefont {Adams},
  \citenamefont {Chesler},\ and\ \citenamefont {Liu}}]{Adams:2013vsa}%
  \BibitemOpen
  \bibfield  {author} {\bibinfo {author} {\bibfnamefont {Allan}\ \bibnamefont
  {Adams}}, \bibinfo {author} {\bibfnamefont {Paul~M.}\ \bibnamefont
  {Chesler}}, \ and\ \bibinfo {author} {\bibfnamefont {Hong}\ \bibnamefont
  {Liu}},\ }\bibfield  {title} {\enquote {\bibinfo {title} {{Holographic
  turbulence}},}\ }\href {\doibase 10.1103/PhysRevLett.112.151602} {\bibfield
  {journal} {\bibinfo  {journal} {Phys.~Rev.~Lett.}\ }\textbf {\bibinfo
  {volume} {112}},\ \bibinfo {pages} {151602} (\bibinfo {year} {2014})},\
  \Eprint {http://arxiv.org/abs/1307.7267} {arXiv:1307.7267 [hep-th]}
  \BibitemShut {NoStop}%
%%CITATION = ARXIV:1307.7267;%%
\bibitem [{\citenamefont {Press}\ and\ \citenamefont
  {Teukolsky}(1972)}]{Press:1972zz}%
  \BibitemOpen
  \bibfield  {author} {\bibinfo {author} {\bibfnamefont {William~H.}\
  \bibnamefont {Press}}\ and\ \bibinfo {author} {\bibfnamefont {Saul~A.}\
  \bibnamefont {Teukolsky}},\ }\bibfield  {title} {\enquote {\bibinfo {title}
  {{Floating Orbits, Superradiant Scattering and the Black-hole Bomb}},}\
  }\href {\doibase 10.1038/238211a0} {\bibfield  {journal} {\bibinfo  {journal}
  {Nature}\ }\textbf {\bibinfo {volume} {238}},\ \bibinfo {pages} {211--212}
  (\bibinfo {year} {1972})}\BibitemShut {NoStop}%
%%CITATION = NATUA,238,211;%%
\bibitem [{\citenamefont {Arvanitaki}\ and\ \citenamefont
  {Dubovsky}(2011)}]{Arvanitaki:2010sy}%
  \BibitemOpen
  \bibfield  {author} {\bibinfo {author} {\bibfnamefont {Asimina}\ \bibnamefont
  {Arvanitaki}}\ and\ \bibinfo {author} {\bibfnamefont {Sergei}\ \bibnamefont
  {Dubovsky}},\ }\bibfield  {title} {\enquote {\bibinfo {title} {{Exploring the
  String Axiverse with Precision Black Hole Physics}},}\ }\href {\doibase
  10.1103/PhysRevD.83.044026} {\bibfield  {journal} {\bibinfo  {journal} {Phys.
  Rev.}\ }\textbf {\bibinfo {volume} {D83}},\ \bibinfo {pages} {044026}
  (\bibinfo {year} {2011})},\ \Eprint {http://arxiv.org/abs/1004.3558}
  {arXiv:1004.3558 [hep-th]} \BibitemShut {NoStop}%
%%CITATION = ARXIV:1004.3558;%%
\bibitem [{\citenamefont {East}(2017)}]{East:2017mrj}%
  \BibitemOpen
  \bibfield  {author} {\bibinfo {author} {\bibfnamefont {William~E.}\
  \bibnamefont {East}},\ }\bibfield  {title} {\enquote {\bibinfo {title}
  {{Superradiant instability of massive vector fields around spinning black
  holes in the relativistic regime}},}\ }\href {\doibase
  10.1103/PhysRevD.96.024004} {\bibfield  {journal} {\bibinfo  {journal} {Phys.
  Rev.}\ }\textbf {\bibinfo {volume} {D96}},\ \bibinfo {pages} {024004}
  (\bibinfo {year} {2017})},\ \Eprint {http://arxiv.org/abs/1705.01544}
  {arXiv:1705.01544 [gr-qc]} \BibitemShut {NoStop}%
%%CITATION = ARXIV:1705.01544;%%
\bibitem [{\citenamefont {Hartnoll}\ \emph {et~al.}(2008)\citenamefont
  {Hartnoll}, \citenamefont {Herzog},\ and\ \citenamefont
  {Horowitz}}]{Hartnoll:2008kx}%
  \BibitemOpen
  \bibfield  {author} {\bibinfo {author} {\bibfnamefont {Sean~A.}\ \bibnamefont
  {Hartnoll}}, \bibinfo {author} {\bibfnamefont {Christopher~P.}\ \bibnamefont
  {Herzog}}, \ and\ \bibinfo {author} {\bibfnamefont {Gary~T.}\ \bibnamefont
  {Horowitz}},\ }\bibfield  {title} {\enquote {\bibinfo {title} {{Holographic
  Superconductors}},}\ }\href {\doibase 10.1088/1126-6708/2008/12/015}
  {\bibfield  {journal} {\bibinfo  {journal} {JHEP}\ }\textbf {\bibinfo
  {volume} {12}},\ \bibinfo {pages} {015} (\bibinfo {year} {2008})},\ \Eprint
  {http://arxiv.org/abs/0810.1563} {arXiv:0810.1563 [hep-th]} \BibitemShut
  {NoStop}%
%%CITATION = ARXIV:0810.1563;%%
\bibitem [{\citenamefont {Gubser}(2008)}]{Gubser:2008px}%
  \BibitemOpen
  \bibfield  {author} {\bibinfo {author} {\bibfnamefont {Steven~S.}\
  \bibnamefont {Gubser}},\ }\bibfield  {title} {\enquote {\bibinfo {title}
  {{Breaking an Abelian gauge symmetry near a black hole horizon}},}\ }\href
  {\doibase 10.1103/PhysRevD.78.065034} {\bibfield  {journal} {\bibinfo
  {journal} {Phys. Rev.}\ }\textbf {\bibinfo {volume} {D78}},\ \bibinfo {pages}
  {065034} (\bibinfo {year} {2008})},\ \Eprint {http://arxiv.org/abs/0801.2977}
  {arXiv:0801.2977 [hep-th]} \BibitemShut {NoStop}%
%%CITATION = ARXIV:0801.2977;%%
\bibitem [{\citenamefont {Bekenstein}(1995)}]{PhysRevD.51.R6608}%
  \BibitemOpen
  \bibfield  {author} {\bibinfo {author} {\bibfnamefont {Jacob~D.}\
  \bibnamefont {Bekenstein}},\ }\bibfield  {title} {\enquote {\bibinfo {title}
  {Novel ``no-scalar-hair'' theorem for black holes},}\ }\href {\doibase
  10.1103/PhysRevD.51.R6608} {\bibfield  {journal} {\bibinfo  {journal} {Phys.
  Rev. D}\ }\textbf {\bibinfo {volume} {51}},\ \bibinfo {pages} {R6608--R6611}
  (\bibinfo {year} {1995})}\BibitemShut {NoStop}%
\bibitem [{\citenamefont {Herdeiro}\ and\ \citenamefont
  {Radu}(2015)}]{Herdeiro:2015waa}%
  \BibitemOpen
  \bibfield  {author} {\bibinfo {author} {\bibfnamefont {Carlos A.~R.}\
  \bibnamefont {Herdeiro}}\ and\ \bibinfo {author} {\bibfnamefont {Eugen}\
  \bibnamefont {Radu}},\ }\bibfield  {title} {\enquote {\bibinfo {title}
  {{Asymptotically flat black holes with scalar hair: a review}},}\ }\bibfield
  {booktitle} {\emph {\bibinfo {booktitle} {{Proceedings, 7th Black Holes
  Workshop 2014: Aveiro, Portugal, December 18-19, 2014}}},\ }\href {\doibase
  10.1142/S0218271815420146} {\bibfield  {journal} {\bibinfo  {journal} {Int.
  J. Mod. Phys.}\ }\textbf {\bibinfo {volume} {D24}},\ \bibinfo {pages}
  {1542014} (\bibinfo {year} {2015})},\ \Eprint
  {http://arxiv.org/abs/1504.08209} {arXiv:1504.08209 [gr-qc]} \BibitemShut
  {NoStop}%
%%CITATION = ARXIV:1504.08209;%%
\bibitem [{\citenamefont {Hollands}\ and\ \citenamefont
  {Wald}(2013)}]{Hollands:2012sf}%
  \BibitemOpen
  \bibfield  {author} {\bibinfo {author} {\bibfnamefont {Stefan}\ \bibnamefont
  {Hollands}}\ and\ \bibinfo {author} {\bibfnamefont {Robert~M.}\ \bibnamefont
  {Wald}},\ }\bibfield  {title} {\enquote {\bibinfo {title} {{Stability of
  Black Holes and Black Branes}},}\ }\href {\doibase 10.1007/s00220-012-1638-1}
  {\bibfield  {journal} {\bibinfo  {journal} {Commun.~Math.~Phys.}\ }\textbf
  {\bibinfo {volume} {321}},\ \bibinfo {pages} {629--680} (\bibinfo {year}
  {2013})},\ \Eprint {http://arxiv.org/abs/1201.0463} {arXiv:1201.0463 [gr-qc]}
  \BibitemShut {NoStop}%
%%CITATION = ARXIV:1201.0463;%%
\bibitem [{\citenamefont {Gregory}\ and\ \citenamefont
  {Laflamme}(1993)}]{Gregory:1993vy}%
  \BibitemOpen
  \bibfield  {author} {\bibinfo {author} {\bibfnamefont {R.}~\bibnamefont
  {Gregory}}\ and\ \bibinfo {author} {\bibfnamefont {R.}~\bibnamefont
  {Laflamme}},\ }\bibfield  {title} {\enquote {\bibinfo {title} {{Black strings
  and p-branes are unstable}},}\ }\href {\doibase 10.1103/PhysRevLett.70.2837}
  {\bibfield  {journal} {\bibinfo  {journal} {Phys. Rev. Lett.}\ }\textbf
  {\bibinfo {volume} {70}},\ \bibinfo {pages} {2837--2840} (\bibinfo {year}
  {1993})},\ \Eprint {http://arxiv.org/abs/hep-th/9301052}
  {arXiv:hep-th/9301052 [hep-th]} \BibitemShut {NoStop}%
%%CITATION = HEP-TH/9301052;%%
\bibitem [{\citenamefont {Lehner}\ and\ \citenamefont
  {Pretorius}(2010)}]{Lehner:2010pn}%
  \BibitemOpen
  \bibfield  {author} {\bibinfo {author} {\bibfnamefont {Luis}\ \bibnamefont
  {Lehner}}\ and\ \bibinfo {author} {\bibfnamefont {Frans}\ \bibnamefont
  {Pretorius}},\ }\bibfield  {title} {\enquote {\bibinfo {title} {{Black
  Strings, Low Viscosity Fluids, and Violation of Cosmic Censorship}},}\ }\href
  {\doibase 10.1103/PhysRevLett.105.101102} {\bibfield  {journal} {\bibinfo
  {journal} {Phys.~Rev.~Lett.}\ }\textbf {\bibinfo {volume} {105}},\ \bibinfo
  {pages} {101102} (\bibinfo {year} {2010})},\ \Eprint
  {http://arxiv.org/abs/1006.5960} {arXiv:1006.5960 [hep-th]} \BibitemShut
  {NoStop}%
%%CITATION = ARXIV:1006.5960;%%
\bibitem [{\citenamefont {Murata}\ \emph {et~al.}(2010)\citenamefont {Murata},
  \citenamefont {Kinoshita},\ and\ \citenamefont {Tanahashi}}]{Murata:2010dx}%
  \BibitemOpen
  \bibfield  {author} {\bibinfo {author} {\bibfnamefont {Keiju}\ \bibnamefont
  {Murata}}, \bibinfo {author} {\bibfnamefont {Shunichiro}\ \bibnamefont
  {Kinoshita}}, \ and\ \bibinfo {author} {\bibfnamefont {Norihiro}\
  \bibnamefont {Tanahashi}},\ }\bibfield  {title} {\enquote {\bibinfo {title}
  {{Non-equilibrium Condensation Process in a Holographic Superconductor}},}\
  }\href {\doibase 10.1007/JHEP07(2010)050} {\bibfield  {journal} {\bibinfo
  {journal} {JHEP}\ }\textbf {\bibinfo {volume} {07}},\ \bibinfo {pages} {050}
  (\bibinfo {year} {2010})},\ \Eprint {http://arxiv.org/abs/1005.0633}
  {arXiv:1005.0633 [hep-th]} \BibitemShut {NoStop}%
%%CITATION = ARXIV:1005.0633;%%
\bibitem [{\citenamefont {Bosch}\ \emph {et~al.}(2016)\citenamefont {Bosch},
  \citenamefont {Green},\ and\ \citenamefont {Lehner}}]{Bosch:2016vcp}%
  \BibitemOpen
  \bibfield  {author} {\bibinfo {author} {\bibfnamefont {Pablo}\ \bibnamefont
  {Bosch}}, \bibinfo {author} {\bibfnamefont {Stephen~R.}\ \bibnamefont
  {Green}}, \ and\ \bibinfo {author} {\bibfnamefont {Luis}\ \bibnamefont
  {Lehner}},\ }\bibfield  {title} {\enquote {\bibinfo {title} {{Nonlinear
  Evolution and Final Fate of Charged Anti--de Sitter Black Hole Superradiant
  Instability}},}\ }\href {\doibase 10.1103/PhysRevLett.116.141102} {\bibfield
  {journal} {\bibinfo  {journal} {Phys.~Rev.~Lett.}\ }\textbf {\bibinfo
  {volume} {116}},\ \bibinfo {pages} {141102} (\bibinfo {year} {2016})},\
  \Eprint {http://arxiv.org/abs/1601.01384} {arXiv:1601.01384 [gr-qc]}
  \BibitemShut {NoStop}%
%%CITATION = ARXIV:1601.01384;%%
\bibitem [{\citenamefont {Dias}\ and\ \citenamefont
  {Masachs}(2017)}]{Dias:2016pma}%
  \BibitemOpen
  \bibfield  {author} {\bibinfo {author} {\bibfnamefont {\'Oscar J.~C.}\
  \bibnamefont {Dias}}\ and\ \bibinfo {author} {\bibfnamefont {Ramon}\
  \bibnamefont {Masachs}},\ }\bibfield  {title} {\enquote {\bibinfo {title}
  {{Hairy black holes and the endpoint of AdS$_4$ charged superradiance}},}\
  }\href {\doibase 10.1007/JHEP02(2017)128} {\bibfield  {journal} {\bibinfo
  {journal} {JHEP}\ }\textbf {\bibinfo {volume} {02}},\ \bibinfo {pages} {128}
  (\bibinfo {year} {2017})},\ \Eprint {http://arxiv.org/abs/1610.03496}
  {arXiv:1610.03496 [hep-th]} \BibitemShut {NoStop}%
%%CITATION = ARXIV:1610.03496;%%
\bibitem [{\citenamefont {Leaver}(1985)}]{Leaver:1985}%
  \BibitemOpen
  \bibfield  {author} {\bibinfo {author} {\bibfnamefont {E.~W.}\ \bibnamefont
  {Leaver}},\ }\bibfield  {title} {\enquote {\bibinfo {title} {An analytic
  representation for the quasi-normal modes of kerr black holes},}\ }\href
  {\doibase 10.1098/rspa.1985.0119} {\bibfield  {journal} {\bibinfo  {journal}
  {Proceedings of the Royal Society A: Mathematical, Physical and Engineering
  Sciences}\ }\textbf {\bibinfo {volume} {402}},\ \bibinfo {pages} {285--298}
  (\bibinfo {year} {1985})}\BibitemShut {NoStop}%
\bibitem [{\citenamefont {Uchikata}\ and\ \citenamefont
  {Yoshida}(2011)}]{Uchikata:2011zz}%
  \BibitemOpen
  \bibfield  {author} {\bibinfo {author} {\bibfnamefont {Nami}\ \bibnamefont
  {Uchikata}}\ and\ \bibinfo {author} {\bibfnamefont {Shijun}\ \bibnamefont
  {Yoshida}},\ }\bibfield  {title} {\enquote {\bibinfo {title} {{Quasinormal
  modes of a massless charged scalar field on a small
  Reissner-Nordstrom-anti-de Sitter black hole}},}\ }\href {\doibase
  10.1103/PhysRevD.83.064020} {\bibfield  {journal} {\bibinfo  {journal} {Phys.
  Rev.}\ }\textbf {\bibinfo {volume} {D83}},\ \bibinfo {pages} {064020}
  (\bibinfo {year} {2011})},\ \Eprint {http://arxiv.org/abs/1109.6737}
  {arXiv:1109.6737 [gr-qc]} \BibitemShut {NoStop}%
%%CITATION = ARXIV:1109.6737;%%
\bibitem [{\citenamefont {Bardeen}\ and\ \citenamefont
  {Horowitz}(1999)}]{Bardeen:1999px}%
  \BibitemOpen
  \bibfield  {author} {\bibinfo {author} {\bibfnamefont {James~M.}\
  \bibnamefont {Bardeen}}\ and\ \bibinfo {author} {\bibfnamefont {Gary~T.}\
  \bibnamefont {Horowitz}},\ }\bibfield  {title} {\enquote {\bibinfo {title}
  {{The Extreme Kerr throat geometry: A Vacuum analog of $\text{AdS}_2 \times
  \text{S}^2$}},}\ }\href {\doibase 10.1103/PhysRevD.60.104030} {\bibfield
  {journal} {\bibinfo  {journal} {Phys.~Rev.}\ }\textbf {\bibinfo {volume}
  {D60}},\ \bibinfo {pages} {104030} (\bibinfo {year} {1999})},\ \Eprint
  {http://arxiv.org/abs/hep-th/9905099} {arXiv:hep-th/9905099 [hep-th]}
  \BibitemShut {NoStop}%
%%CITATION = HEP-TH/9905099;%%
\bibitem [{\citenamefont {Breitenlohner}\ and\ \citenamefont
  {Freedman}(1982)}]{Breitenlohner:1982bm}%
  \BibitemOpen
  \bibfield  {author} {\bibinfo {author} {\bibfnamefont {Peter}\ \bibnamefont
  {Breitenlohner}}\ and\ \bibinfo {author} {\bibfnamefont {Daniel~Z.}\
  \bibnamefont {Freedman}},\ }\bibfield  {title} {\enquote {\bibinfo {title}
  {{Positive Energy in anti-De Sitter Backgrounds and Gauged Extended
  Supergravity}},}\ }\href {\doibase 10.1016/0370-2693(82)90643-8} {\bibfield
  {journal} {\bibinfo  {journal} {Phys. Lett.}\ }\textbf {\bibinfo {volume}
  {115B}},\ \bibinfo {pages} {197--201} (\bibinfo {year} {1982})}\BibitemShut
  {NoStop}%
%%CITATION = PHLTA,115B,197;%%
\bibitem [{\citenamefont {Abdalla}\ \emph {et~al.}(2010)\citenamefont
  {Abdalla}, \citenamefont {Pellicer}, \citenamefont {de~Oliveira},\ and\
  \citenamefont {Pavan}}]{PhysRevD.82.124033}%
  \BibitemOpen
  \bibfield  {author} {\bibinfo {author} {\bibfnamefont {E.}~\bibnamefont
  {Abdalla}}, \bibinfo {author} {\bibfnamefont {C.~E.}\ \bibnamefont
  {Pellicer}}, \bibinfo {author} {\bibfnamefont {Jeferson}\ \bibnamefont
  {de~Oliveira}}, \ and\ \bibinfo {author} {\bibfnamefont {A.~B.}\ \bibnamefont
  {Pavan}},\ }\bibfield  {title} {\enquote {\bibinfo {title} {Phase transitions
  and regions of stability in reissner-nordstr\"om holographic
  superconductors},}\ }\href {\doibase 10.1103/PhysRevD.82.124033} {\bibfield
  {journal} {\bibinfo  {journal} {Phys. Rev. D}\ }\textbf {\bibinfo {volume}
  {82}},\ \bibinfo {pages} {124033} (\bibinfo {year} {2010})}\BibitemShut
  {NoStop}%
\bibitem [{\citenamefont {Maeda}\ \emph {et~al.}(2010)\citenamefont {Maeda},
  \citenamefont {Fujii},\ and\ \citenamefont {Koga}}]{PhysRevD.81.124020}%
  \BibitemOpen
  \bibfield  {author} {\bibinfo {author} {\bibfnamefont {Kengo}\ \bibnamefont
  {Maeda}}, \bibinfo {author} {\bibfnamefont {Shunsuke}\ \bibnamefont {Fujii}},
  \ and\ \bibinfo {author} {\bibfnamefont {Jun-ichirou}\ \bibnamefont {Koga}},\
  }\bibfield  {title} {\enquote {\bibinfo {title} {Final fate of instability of
  reissner-nordstr\"om-anti-de sitter black holes by charged complex scalar
  fields},}\ }\href {\doibase 10.1103/PhysRevD.81.124020} {\bibfield  {journal}
  {\bibinfo  {journal} {Phys. Rev. D}\ }\textbf {\bibinfo {volume} {81}},\
  \bibinfo {pages} {124020} (\bibinfo {year} {2010})}\BibitemShut {NoStop}%
\bibitem [{\citenamefont {Hollands}\ and\ \citenamefont
  {Ishibashi}(2015)}]{Hollands:2014lra}%
  \BibitemOpen
  \bibfield  {author} {\bibinfo {author} {\bibfnamefont {Stefan}\ \bibnamefont
  {Hollands}}\ and\ \bibinfo {author} {\bibfnamefont {Akihiro}\ \bibnamefont
  {Ishibashi}},\ }\bibfield  {title} {\enquote {\bibinfo {title}
  {{Instabilities of extremal rotating black holes in higher dimensions}},}\
  }\href {\doibase 10.1007/s00220-015-2410-0} {\bibfield  {journal} {\bibinfo
  {journal} {Commun. Math. Phys.}\ }\textbf {\bibinfo {volume} {339}},\
  \bibinfo {pages} {949--1002} (\bibinfo {year} {2015})},\ \Eprint
  {http://arxiv.org/abs/1408.0801} {arXiv:1408.0801 [hep-th]} \BibitemShut
  {NoStop}%
%%CITATION = ARXIV:1408.0801;%%
\bibitem [{\citenamefont {Zimmerman}(2017)}]{Zimmerman:2016qtn}%
  \BibitemOpen
  \bibfield  {author} {\bibinfo {author} {\bibfnamefont {Peter}\ \bibnamefont
  {Zimmerman}},\ }\bibfield  {title} {\enquote {\bibinfo {title} {{Horizon
  instability of extremal Reissner-Nordstr\"om black holes to charged
  perturbations}},}\ }\href {\doibase 10.1103/PhysRevD.95.124032} {\bibfield
  {journal} {\bibinfo  {journal} {Phys. Rev.}\ }\textbf {\bibinfo {volume}
  {D95}},\ \bibinfo {pages} {124032} (\bibinfo {year} {2017})},\ \Eprint
  {http://arxiv.org/abs/1612.03172} {arXiv:1612.03172 [gr-qc]} \BibitemShut
  {NoStop}%
%%CITATION = ARXIV:1612.03172;%%
\bibitem [{\citenamefont {Aretakis}(2015)}]{Aretakis:2012ei}%
  \BibitemOpen
  \bibfield  {author} {\bibinfo {author} {\bibfnamefont {Stefanos}\
  \bibnamefont {Aretakis}},\ }\bibfield  {title} {\enquote {\bibinfo {title}
  {{Horizon Instability of Extremal Black Holes}},}\ }\href {\doibase
  10.4310/ATMP.2015.v19.n3.a1} {\bibfield  {journal} {\bibinfo  {journal} {Adv.
  Theor. Math. Phys.}\ }\textbf {\bibinfo {volume} {19}},\ \bibinfo {pages}
  {507--530} (\bibinfo {year} {2015})},\ \Eprint
  {http://arxiv.org/abs/1206.6598} {arXiv:1206.6598 [gr-qc]} \BibitemShut
  {NoStop}%
%%CITATION = ARXIV:1206.6598;%%
\bibitem [{\citenamefont {Wald}(1984)}]{Wald:1984}%
  \BibitemOpen
  \bibfield  {author} {\bibinfo {author} {\bibfnamefont {Robert~M.}\
  \bibnamefont {Wald}},\ }\href@noop {} {\emph {\bibinfo {title} {General
  Relativity}}}\ (\bibinfo  {publisher} {University of Chicago Press},\
  \bibinfo {address} {Chicago, IL},\ \bibinfo {year} {1984})\BibitemShut
  {NoStop}%
\bibitem [{\citenamefont {Dias}\ \emph {et~al.}(2010)\citenamefont {Dias},
  \citenamefont {Figueras}, \citenamefont {Monteiro}, \citenamefont {Reall},\
  and\ \citenamefont {Santos}}]{Dias:2010eu}%
  \BibitemOpen
  \bibfield  {author} {\bibinfo {author} {\bibfnamefont {Oscar J.~C.}\
  \bibnamefont {Dias}}, \bibinfo {author} {\bibfnamefont {Pau}\ \bibnamefont
  {Figueras}}, \bibinfo {author} {\bibfnamefont {Ricardo}\ \bibnamefont
  {Monteiro}}, \bibinfo {author} {\bibfnamefont {Harvey~S.}\ \bibnamefont
  {Reall}}, \ and\ \bibinfo {author} {\bibfnamefont {Jorge~E.}\ \bibnamefont
  {Santos}},\ }\bibfield  {title} {\enquote {\bibinfo {title} {{An instability
  of higher-dimensional rotating black holes}},}\ }\href {\doibase
  10.1007/JHEP05(2010)076} {\bibfield  {journal} {\bibinfo  {journal} {JHEP}\
  }\textbf {\bibinfo {volume} {05}},\ \bibinfo {pages} {076} (\bibinfo {year}
  {2010})},\ \Eprint {http://arxiv.org/abs/1001.4527} {arXiv:1001.4527
  [hep-th]} \BibitemShut {NoStop}%
%%CITATION = ARXIV:1001.4527;%%
\bibitem [{\citenamefont {Bekenstein}(1973)}]{PhysRevD.7.949}%
  \BibitemOpen
  \bibfield  {author} {\bibinfo {author} {\bibfnamefont {Jacob~D.}\
  \bibnamefont {Bekenstein}},\ }\bibfield  {title} {\enquote {\bibinfo {title}
  {Extraction of energy and charge from a black hole},}\ }\href {\doibase
  10.1103/PhysRevD.7.949} {\bibfield  {journal} {\bibinfo  {journal} {Phys.
  Rev. D}\ }\textbf {\bibinfo {volume} {7}},\ \bibinfo {pages} {949--953}
  (\bibinfo {year} {1973})}\BibitemShut {NoStop}%
\bibitem [{\citenamefont {Gautschi}(1967)}]{Gautschi:1967}%
  \BibitemOpen
  \bibfield  {author} {\bibinfo {author} {\bibfnamefont {Walter}\ \bibnamefont
  {Gautschi}},\ }\bibfield  {title} {\enquote {\bibinfo {title} {Computational
  aspects of three-term recurrence relations},}\ }\href {\doibase
  10.1137/1009002} {\bibfield  {journal} {\bibinfo  {journal} {{SIAM} Review}\
  }\textbf {\bibinfo {volume} {9}},\ \bibinfo {pages} {24--82} (\bibinfo {year}
  {1967})}\BibitemShut {NoStop}%
\bibitem [{\citenamefont {Richartz}\ and\ \citenamefont
  {Giugno}(2014)}]{Richartz:2014}%
  \BibitemOpen
  \bibfield  {author} {\bibinfo {author} {\bibfnamefont {Maur{\'{\i}}cio}\
  \bibnamefont {Richartz}}\ and\ \bibinfo {author} {\bibfnamefont {Davi}\
  \bibnamefont {Giugno}},\ }\bibfield  {title} {\enquote {\bibinfo {title}
  {Quasinormal modes of charged fields around a reissner-nordstr\"{o}m black
  hole},}\ }\href {\doibase 10.1103/physrevd.90.124011} {\bibfield  {journal}
  {\bibinfo  {journal} {Physical Review D}\ }\textbf {\bibinfo {volume} {90}}
  (\bibinfo {year} {2014}),\ 10.1103/physrevd.90.124011}\BibitemShut {NoStop}%
\bibitem [{\citenamefont {Chesler}\ and\ \citenamefont
  {Yaffe}(2014)}]{Chesler:2013lia}%
  \BibitemOpen
  \bibfield  {author} {\bibinfo {author} {\bibfnamefont {Paul~M.}\ \bibnamefont
  {Chesler}}\ and\ \bibinfo {author} {\bibfnamefont {Laurence~G.}\ \bibnamefont
  {Yaffe}},\ }\bibfield  {title} {\enquote {\bibinfo {title} {{Numerical
  solution of gravitational dynamics in asymptotically anti-de Sitter
  spacetimes}},}\ }\href {\doibase 10.1007/JHEP07(2014)086} {\bibfield
  {journal} {\bibinfo  {journal} {JHEP}\ }\textbf {\bibinfo {volume} {07}},\
  \bibinfo {pages} {086} (\bibinfo {year} {2014})},\ \Eprint
  {http://arxiv.org/abs/1309.1439} {arXiv:1309.1439 [hep-th]} \BibitemShut
  {NoStop}%
%%CITATION = ARXIV:1309.1439;%%
\bibitem [{\citenamefont {Calabrese}\ \emph {et~al.}(2004)\citenamefont
  {Calabrese}, \citenamefont {Lehner}, \citenamefont {Reula}, \citenamefont
  {Sarbach},\ and\ \citenamefont {Tiglio}}]{Calabrese:2003vx}%
  \BibitemOpen
  \bibfield  {author} {\bibinfo {author} {\bibfnamefont {Gioel}\ \bibnamefont
  {Calabrese}}, \bibinfo {author} {\bibfnamefont {Luis}\ \bibnamefont
  {Lehner}}, \bibinfo {author} {\bibfnamefont {Oscar}\ \bibnamefont {Reula}},
  \bibinfo {author} {\bibfnamefont {Olivier}\ \bibnamefont {Sarbach}}, \ and\
  \bibinfo {author} {\bibfnamefont {Manuel}\ \bibnamefont {Tiglio}},\
  }\bibfield  {title} {\enquote {\bibinfo {title} {{Summation by parts and
  dissipation for domains with excised regions}},}\ }\href {\doibase
  10.1088/0264-9381/21/24/004} {\bibfield  {journal} {\bibinfo  {journal}
  {Class. Quant. Grav.}\ }\textbf {\bibinfo {volume} {21}},\ \bibinfo {pages}
  {5735--5758} (\bibinfo {year} {2004})},\ \Eprint
  {http://arxiv.org/abs/gr-qc/0308007} {arXiv:gr-qc/0308007 [gr-qc]}
  \BibitemShut {NoStop}%
%%CITATION = GR-QC/0308007;%%
\bibitem [{\citenamefont {Calabrese}\ \emph {et~al.}(2003)\citenamefont
  {Calabrese}, \citenamefont {Lehner}, \citenamefont {Neilsen}, \citenamefont
  {Pullin}, \citenamefont {Reula}, \citenamefont {Sarbach},\ and\ \citenamefont
  {Tiglio}}]{Calabrese:2003yd}%
  \BibitemOpen
  \bibfield  {author} {\bibinfo {author} {\bibfnamefont {Gioel}\ \bibnamefont
  {Calabrese}}, \bibinfo {author} {\bibfnamefont {Luis}\ \bibnamefont
  {Lehner}}, \bibinfo {author} {\bibfnamefont {David}\ \bibnamefont {Neilsen}},
  \bibinfo {author} {\bibfnamefont {Jorge}\ \bibnamefont {Pullin}}, \bibinfo
  {author} {\bibfnamefont {Oscar}\ \bibnamefont {Reula}}, \bibinfo {author}
  {\bibfnamefont {Olivier}\ \bibnamefont {Sarbach}}, \ and\ \bibinfo {author}
  {\bibfnamefont {Manuel}\ \bibnamefont {Tiglio}},\ }\bibfield  {title}
  {\enquote {\bibinfo {title} {{Novel finite differencing techniques for
  numerical relativity: Application to black hole excision}},}\ }\href
  {\doibase 10.1088/0264-9381/20/20/102} {\bibfield  {journal} {\bibinfo
  {journal} {Class. Quant. Grav.}\ }\textbf {\bibinfo {volume} {20}},\ \bibinfo
  {pages} {L245--L252} (\bibinfo {year} {2003})},\ \Eprint
  {http://arxiv.org/abs/gr-qc/0302072} {arXiv:gr-qc/0302072 [gr-qc]}
  \BibitemShut {NoStop}%
%%CITATION = GR-QC/0302072;%%
\bibitem [{\citenamefont {Bosch}\ \emph {et~al.}(2017)\citenamefont {Bosch},
  \citenamefont {Buchel},\ and\ \citenamefont {Lehner}}]{Bosch:2017ccw}%
  \BibitemOpen
  \bibfield  {author} {\bibinfo {author} {\bibfnamefont {Pablo}\ \bibnamefont
  {Bosch}}, \bibinfo {author} {\bibfnamefont {Alex}\ \bibnamefont {Buchel}}, \
  and\ \bibinfo {author} {\bibfnamefont {Luis}\ \bibnamefont {Lehner}},\
  }\bibfield  {title} {\enquote {\bibinfo {title} {{Unstable horizons and
  singularity development in holography}},}\ }\href {\doibase
  10.1007/JHEP07(2017)135} {\bibfield  {journal} {\bibinfo  {journal} {JHEP}\
  }\textbf {\bibinfo {volume} {07}},\ \bibinfo {pages} {135} (\bibinfo {year}
  {2017})},\ \Eprint {http://arxiv.org/abs/1704.05454} {arXiv:1704.05454
  [hep-th]} \BibitemShut {NoStop}%
%%CITATION = ARXIV:1704.05454;%%
\bibitem [{\citenamefont {Siemonsen}\ and\ \citenamefont
  {East}(2019)}]{Siemonsen:2019ebd}%
  \BibitemOpen
  \bibfield  {author} {\bibinfo {author} {\bibfnamefont {Nils}\ \bibnamefont
  {Siemonsen}}\ and\ \bibinfo {author} {\bibfnamefont {William~E.}\
  \bibnamefont {East}},\ }\bibfield  {title} {\enquote {\bibinfo {title}
  {{Gravitational wave signatures of ultralight vector bosons from black hole
  superradiance}},}\ }\href@noop {} {\  (\bibinfo {year} {2019})},\ \Eprint
  {http://arxiv.org/abs/1910.09476} {arXiv:1910.09476 [gr-qc]} \BibitemShut
  {NoStop}%
%%CITATION = ARXIV:1910.09476;%%
\bibitem [{\citenamefont {Arvanitaki}\ \emph {et~al.}(2015)\citenamefont
  {Arvanitaki}, \citenamefont {Baryakhtar},\ and\ \citenamefont
  {Huang}}]{Arvanitaki:2014wva}%
  \BibitemOpen
  \bibfield  {author} {\bibinfo {author} {\bibfnamefont {Asimina}\ \bibnamefont
  {Arvanitaki}}, \bibinfo {author} {\bibfnamefont {Masha}\ \bibnamefont
  {Baryakhtar}}, \ and\ \bibinfo {author} {\bibfnamefont {Xinlu}\ \bibnamefont
  {Huang}},\ }\bibfield  {title} {\enquote {\bibinfo {title} {{Discovering the
  QCD Axion with Black Holes and Gravitational Waves}},}\ }\href {\doibase
  10.1103/PhysRevD.91.084011} {\bibfield  {journal} {\bibinfo  {journal}
  {Phys.~Rev.}\ }\textbf {\bibinfo {volume} {D91}},\ \bibinfo {pages} {084011}
  (\bibinfo {year} {2015})},\ \Eprint {http://arxiv.org/abs/1411.2263}
  {arXiv:1411.2263 [hep-ph]} \BibitemShut {NoStop}%
%%CITATION = ARXIV:1411.2263;%%
\bibitem [{\citenamefont {Chesler}\ and\ \citenamefont
  {Lowe}(2019)}]{Chesler:2018txn}%
  \BibitemOpen
  \bibfield  {author} {\bibinfo {author} {\bibfnamefont {Paul~M.}\ \bibnamefont
  {Chesler}}\ and\ \bibinfo {author} {\bibfnamefont {David~A.}\ \bibnamefont
  {Lowe}},\ }\bibfield  {title} {\enquote {\bibinfo {title} {{Nonlinear
  Evolution of the AdS$_4$ Superradiant Instability}},}\ }\href {\doibase
  10.1103/PhysRevLett.122.181101} {\bibfield  {journal} {\bibinfo  {journal}
  {Phys. Rev. Lett.}\ }\textbf {\bibinfo {volume} {122}},\ \bibinfo {pages}
  {181101} (\bibinfo {year} {2019})},\ \Eprint
  {http://arxiv.org/abs/1801.09711} {arXiv:1801.09711 [gr-qc]} \BibitemShut
  {NoStop}%
%%CITATION = ARXIV:1801.09711;%%
\bibitem [{\citenamefont {Dias}\ \emph {et~al.}(2015)\citenamefont {Dias},
  \citenamefont {Santos},\ and\ \citenamefont {Way}}]{Dias:2015rxy}%
  \BibitemOpen
  \bibfield  {author} {\bibinfo {author} {\bibfnamefont {Óscar J.~C.}\
  \bibnamefont {Dias}}, \bibinfo {author} {\bibfnamefont {Jorge~E.}\
  \bibnamefont {Santos}}, \ and\ \bibinfo {author} {\bibfnamefont {Benson}\
  \bibnamefont {Way}},\ }\bibfield  {title} {\enquote {\bibinfo {title} {{Black
  holes with a single Killing vector field: black resonators}},}\ }\href
  {\doibase 10.1007/JHEP12(2015)171} {\bibfield  {journal} {\bibinfo  {journal}
  {JHEP}\ }\textbf {\bibinfo {volume} {12}},\ \bibinfo {pages} {171} (\bibinfo
  {year} {2015})},\ \Eprint {http://arxiv.org/abs/1505.04793} {arXiv:1505.04793
  [hep-th]} \BibitemShut {NoStop}%
%%CITATION = ARXIV:1505.04793;%%
\bibitem [{\citenamefont {Green}\ \emph {et~al.}(2016)\citenamefont {Green},
  \citenamefont {Hollands}, \citenamefont {Ishibashi},\ and\ \citenamefont
  {Wald}}]{Green:2015kur}%
  \BibitemOpen
  \bibfield  {author} {\bibinfo {author} {\bibfnamefont {Stephen~R.}\
  \bibnamefont {Green}}, \bibinfo {author} {\bibfnamefont {Stefan}\
  \bibnamefont {Hollands}}, \bibinfo {author} {\bibfnamefont {Akihiro}\
  \bibnamefont {Ishibashi}}, \ and\ \bibinfo {author} {\bibfnamefont
  {Robert~M.}\ \bibnamefont {Wald}},\ }\bibfield  {title} {\enquote {\bibinfo
  {title} {{Superradiant instabilities of asymptotically anti-de Sitter black
  holes}},}\ }\href {\doibase 10.1088/0264-9381/33/12/125022} {\bibfield
  {journal} {\bibinfo  {journal} {Class.~Quant.~Grav.}\ }\textbf {\bibinfo
  {volume} {33}},\ \bibinfo {pages} {125022} (\bibinfo {year} {2016})},\
  \Eprint {http://arxiv.org/abs/1512.02644} {arXiv:1512.02644 [gr-qc]}
  \BibitemShut {NoStop}%
%%CITATION = ARXIV:1512.02644;%%
\bibitem [{\citenamefont {Horowitz}\ \emph {et~al.}(2011)\citenamefont
  {Horowitz}, \citenamefont {Santos},\ and\ \citenamefont
  {Way}}]{Horowitz:2011dz}%
  \BibitemOpen
  \bibfield  {author} {\bibinfo {author} {\bibfnamefont {Gary~T.}\ \bibnamefont
  {Horowitz}}, \bibinfo {author} {\bibfnamefont {Jorge~E.}\ \bibnamefont
  {Santos}}, \ and\ \bibinfo {author} {\bibfnamefont {Benson}\ \bibnamefont
  {Way}},\ }\bibfield  {title} {\enquote {\bibinfo {title} {{A Holographic
  Josephson Junction}},}\ }\href {\doibase 10.1103/PhysRevLett.106.221601}
  {\bibfield  {journal} {\bibinfo  {journal} {Phys. Rev. Lett.}\ }\textbf
  {\bibinfo {volume} {106}},\ \bibinfo {pages} {221601} (\bibinfo {year}
  {2011})},\ \Eprint {http://arxiv.org/abs/1101.3326} {arXiv:1101.3326
  [hep-th]} \BibitemShut {NoStop}%
%%CITATION = ARXIV:1101.3326;%%
\bibitem [{\citenamefont {Giesler}\ \emph {et~al.}(2019)\citenamefont
  {Giesler}, \citenamefont {Isi}, \citenamefont {Scheel},\ and\ \citenamefont
  {Teukolsky}}]{Giesler:2019uxc}%
  \BibitemOpen
  \bibfield  {author} {\bibinfo {author} {\bibfnamefont {Matthew}\ \bibnamefont
  {Giesler}}, \bibinfo {author} {\bibfnamefont {Maximiliano}\ \bibnamefont
  {Isi}}, \bibinfo {author} {\bibfnamefont {Mark}\ \bibnamefont {Scheel}}, \
  and\ \bibinfo {author} {\bibfnamefont {Saul}\ \bibnamefont {Teukolsky}},\
  }\bibfield  {title} {\enquote {\bibinfo {title} {{Black hole ringdown: the
  importance of overtones}},}\ }\href@noop {} {\  (\bibinfo {year} {2019})},\
  \Eprint {http://arxiv.org/abs/1903.08284} {arXiv:1903.08284 [gr-qc]}
  \BibitemShut {NoStop}%
%%CITATION = ARXIV:1903.08284;%%
\bibitem [{\citenamefont {Isi}\ \emph {et~al.}(2019)\citenamefont {Isi},
  \citenamefont {Giesler}, \citenamefont {Farr}, \citenamefont {Scheel},\ and\
  \citenamefont {Teukolsky}}]{Isi:2019aib}%
  \BibitemOpen
  \bibfield  {author} {\bibinfo {author} {\bibfnamefont {Maximiliano}\
  \bibnamefont {Isi}}, \bibinfo {author} {\bibfnamefont {Matthew}\ \bibnamefont
  {Giesler}}, \bibinfo {author} {\bibfnamefont {Will~M.}\ \bibnamefont {Farr}},
  \bibinfo {author} {\bibfnamefont {Mark~A.}\ \bibnamefont {Scheel}}, \ and\
  \bibinfo {author} {\bibfnamefont {Saul~A.}\ \bibnamefont {Teukolsky}},\
  }\bibfield  {title} {\enquote {\bibinfo {title} {{Testing the no-hair theorem
  with GW150914}},}\ }\href {\doibase 10.1103/PhysRevLett.123.111102}
  {\bibfield  {journal} {\bibinfo  {journal} {Phys. Rev. Lett.}\ }\textbf
  {\bibinfo {volume} {123}},\ \bibinfo {pages} {111102} (\bibinfo {year}
  {2019})},\ \Eprint {http://arxiv.org/abs/1905.00869} {arXiv:1905.00869
  [gr-qc]} \BibitemShut {NoStop}%
%%CITATION = ARXIV:1905.00869;%%
\bibitem [{\citenamefont {Ota}\ and\ \citenamefont
  {Chirenti}(2019)}]{Ota:2019bzl}%
  \BibitemOpen
  \bibfield  {author} {\bibinfo {author} {\bibfnamefont {Iara}\ \bibnamefont
  {Ota}}\ and\ \bibinfo {author} {\bibfnamefont {Cecilia}\ \bibnamefont
  {Chirenti}},\ }\bibfield  {title} {\enquote {\bibinfo {title} {{Overtones or
  higher harmonics? Prospects for testing the no-hair theorem with
  gravitational wave detections}},}\ }\href@noop {} {\  (\bibinfo {year}
  {2019})},\ \Eprint {http://arxiv.org/abs/1911.00440} {arXiv:1911.00440
  [gr-qc]} \BibitemShut {NoStop}%
%%CITATION = ARXIV:1911.00440;%%
\bibitem [{\citenamefont {Zlochower}\ \emph {et~al.}(2003)\citenamefont
  {Zlochower}, \citenamefont {Gomez}, \citenamefont {Husa}, \citenamefont
  {Lehner},\ and\ \citenamefont {Winicour}}]{Zlochower:2003yh}%
  \BibitemOpen
  \bibfield  {author} {\bibinfo {author} {\bibfnamefont {Yosef}\ \bibnamefont
  {Zlochower}}, \bibinfo {author} {\bibfnamefont {Roberto}\ \bibnamefont
  {Gomez}}, \bibinfo {author} {\bibfnamefont {Sascha}\ \bibnamefont {Husa}},
  \bibinfo {author} {\bibfnamefont {Luis}\ \bibnamefont {Lehner}}, \ and\
  \bibinfo {author} {\bibfnamefont {Jeffrey}\ \bibnamefont {Winicour}},\
  }\bibfield  {title} {\enquote {\bibinfo {title} {{Mode coupling in the
  nonlinear response of black holes}},}\ }\href {\doibase
  10.1103/PhysRevD.68.084014} {\bibfield  {journal} {\bibinfo  {journal} {Phys.
  Rev.}\ }\textbf {\bibinfo {volume} {D68}},\ \bibinfo {pages} {084014}
  (\bibinfo {year} {2003})},\ \Eprint {http://arxiv.org/abs/gr-qc/0306098}
  {arXiv:gr-qc/0306098 [gr-qc]} \BibitemShut {NoStop}%
%%CITATION = GR-QC/0306098;%%
\bibitem [{\citenamefont {East}\ \emph {et~al.}(2014)\citenamefont {East},
  \citenamefont {Ramazanoğlu},\ and\ \citenamefont
  {Pretorius}}]{East:2013mfa}%
  \BibitemOpen
  \bibfield  {author} {\bibinfo {author} {\bibfnamefont {William~E.}\
  \bibnamefont {East}}, \bibinfo {author} {\bibfnamefont {Fethi~M.}\
  \bibnamefont {Ramazanoğlu}}, \ and\ \bibinfo {author} {\bibfnamefont
  {Frans}\ \bibnamefont {Pretorius}},\ }\bibfield  {title} {\enquote {\bibinfo
  {title} {{Black Hole Superradiance in Dynamical Spacetime}},}\ }\href
  {\doibase 10.1103/PhysRevD.89.061503} {\bibfield  {journal} {\bibinfo
  {journal} {Phys. Rev.}\ }\textbf {\bibinfo {volume} {D89}},\ \bibinfo {pages}
  {061503} (\bibinfo {year} {2014})},\ \Eprint {http://arxiv.org/abs/1312.4529}
  {arXiv:1312.4529 [gr-qc]} \BibitemShut {NoStop}%
%%CITATION = ARXIV:1312.4529;%%
\bibitem [{\citenamefont {Yang}\ \emph
  {et~al.}(2015{\natexlab{b}})\citenamefont {Yang}, \citenamefont {Zhang},
  \citenamefont {Green},\ and\ \citenamefont {Lehner}}]{Yang:2015jja}%
  \BibitemOpen
  \bibfield  {author} {\bibinfo {author} {\bibfnamefont {Huan}\ \bibnamefont
  {Yang}}, \bibinfo {author} {\bibfnamefont {Fan}\ \bibnamefont {Zhang}},
  \bibinfo {author} {\bibfnamefont {Stephen~R.}\ \bibnamefont {Green}}, \ and\
  \bibinfo {author} {\bibfnamefont {Luis}\ \bibnamefont {Lehner}},\ }\bibfield
  {title} {\enquote {\bibinfo {title} {{Coupled Oscillator Model for Nonlinear
  Gravitational Perturbations}},}\ }\href {\doibase 10.1103/PhysRevD.91.084007}
  {\bibfield  {journal} {\bibinfo  {journal} {Phys.~Rev.}\ }\textbf {\bibinfo
  {volume} {D91}},\ \bibinfo {pages} {084007} (\bibinfo {year}
  {2015}{\natexlab{b}})},\ \Eprint {http://arxiv.org/abs/1502.08051}
  {arXiv:1502.08051 [gr-qc]} \BibitemShut {NoStop}%
%%CITATION = ARXIV:1502.08051;%%
\bibitem [{\citenamefont {Horowitz}\ and\ \citenamefont
  {Hubeny}(2000)}]{Horowitz:1999jd}%
  \BibitemOpen
  \bibfield  {author} {\bibinfo {author} {\bibfnamefont {Gary~T.}\ \bibnamefont
  {Horowitz}}\ and\ \bibinfo {author} {\bibfnamefont {Veronika~E.}\
  \bibnamefont {Hubeny}},\ }\bibfield  {title} {\enquote {\bibinfo {title}
  {{Quasinormal modes of AdS black holes and the approach to thermal
  equilibrium}},}\ }\href {\doibase 10.1103/PhysRevD.62.024027} {\bibfield
  {journal} {\bibinfo  {journal} {Phys. Rev.}\ }\textbf {\bibinfo {volume}
  {D62}},\ \bibinfo {pages} {024027} (\bibinfo {year} {2000})},\ \Eprint
  {http://arxiv.org/abs/hep-th/9909056} {arXiv:hep-th/9909056 [hep-th]}
  \BibitemShut {NoStop}%
%%CITATION = HEP-TH/9909056;%%
\bibitem [{\citenamefont {Berti}\ and\ \citenamefont
  {Kokkotas}(2003)}]{Berti:2003ud}%
  \BibitemOpen
  \bibfield  {author} {\bibinfo {author} {\bibfnamefont {E.}~\bibnamefont
  {Berti}}\ and\ \bibinfo {author} {\bibfnamefont {K.~D.}\ \bibnamefont
  {Kokkotas}},\ }\bibfield  {title} {\enquote {\bibinfo {title} {{Quasinormal
  modes of Reissner-Nordstrom-anti-de Sitter black holes: Scalar,
  electromagnetic and gravitational perturbations}},}\ }\href {\doibase
  10.1103/PhysRevD.67.064020} {\bibfield  {journal} {\bibinfo  {journal} {Phys.
  Rev.}\ }\textbf {\bibinfo {volume} {D67}},\ \bibinfo {pages} {064020}
  (\bibinfo {year} {2003})},\ \Eprint {http://arxiv.org/abs/gr-qc/0301052}
  {arXiv:gr-qc/0301052 [gr-qc]} \BibitemShut {NoStop}%
%%CITATION = GR-QC/0301052;%%
\end{thebibliography}%

\end{document}